\documentclass[%
 reprint,
superscriptaddress,
showpacs,preprintnumbers,
nofootinbib,
 amsmath,amssymb,
 aps,
]{revtex4}

\usepackage{graphicx}
\usepackage{dcolumn}
\usepackage{bm,braket}
\usepackage{url}

\input{colordvi.tex}
\begin{document}

\preprint{-}

\title{Computation approach for CMB bispectrum from primordial magnetic fields}
\author{Maresuke Shiraishi}
\email{mare@a.phys.nagoya-u.ac.jp}
\affiliation{Department of Physics and Astrophysics, Nagoya University,
Aichi 464-8602, Japan}
\author{Daisuke Nitta}
\affiliation{Department of Physics and Astrophysics, Nagoya University,
Aichi 464-8602, Japan}
\author{Shuichiro Yokoyama}
\affiliation{Department of Physics and Astrophysics, Nagoya University,
Aichi 464-8602, Japan}
\author{Kiyotomo Ichiki}
\affiliation{Department of Physics and Astrophysics, Nagoya University,
Aichi 464-8602, Japan}
\author{Keitaro Takahashi}
\affiliation{Graduate School of Science and Technology, Kumamoto University, 2-39-1 Kurokami, Kumamoto 860-8555, Japan}

\date{\today}

\begin{abstract}
We present a detailed calculation of our previous short
paper~[M. Shiraishi, D. Nitta, S. Yokoyama, K. Ichiki, and K. Takahashi, Phys. Rev. D 82, 121302 (2010).] in which we have investigated a constraint
on the magnetic field strength through comic microwave background 
temperature bispectrum of vector modes induced from primordial magnetic
fields.  By taking into account full angular dependence of the
bispectrum with spin spherical harmonics and Wigner symbols, we
explicitly show that the cosmic microwave background bispectrum induced from the
statistical-isotropic primordial vector fluctuations can be also
described as an angle-averaged form in the rotationally invariant way.
We also study the cases with different spectral indices of the power
spectrum of the primordial magnetic fields.

\end{abstract}

\pacs{98.80.Cq, 98.62.En, 98.70.Vc}
\maketitle


\section{Introduction}\label{sec:intro}


Recent observational consequences have shown the existence of ${\cal
O}(10^{-6}) {\rm G}$ magnetic fields in galaxies and clusters of
galaxies at redshift $z \sim 0.7 - 2.0$ \cite{Bernet:2008qp,
Wolfe:2008nk, Kronberg:2007dy}. One of the scenarios to realize this is
an amplification of the magnetic fields by the galactic dynamo mechanism
(e.g. \cite{Widrow:2002ud}), which requires ${\cal O}(10^{-20}) {\rm G}$
seed fields prior to the galaxy formation.  A variety of studies have
suggested the possibility of generating the seed fields at the
inflationary epoch \cite{Martin:2007ue, Bamba:2006ga}, the cosmic phase
transitions \cite{Stevens:2010ym, Kahniashvili:2009qi}, and cosmological
recombination \cite{Ichiki:2006cd, Maeda:2008dv, Fenu:2010kh} 
and also there have been many studies about the constraint on
the strength of primordial magnetic fields (PMFs) through the impact
on the cosmic microwave background (CMB) anisotropies, in particular,
the CMB power spectrum sourced from the PMFs~\cite{Subramanian:1998fn, Durrer:1998ya, Mack:2001gc, Lewis:2004ef, Paoletti:2008ck, Shaw:2009nf}.

Recently, in
Refs.~\cite{Brown:2005kr,Seshadri:2009sy,Caprini:2009vk,Cai:2010uw,
Trivedi:2010gi}, the authors investigated the contribution to the
bispectrum of the CMB temperature fluctuations from the scalar mode PMFs
and roughly estimated the limit on the amplitude of the PMFs.
Because the temperature fluctuations induced by the PMFs have the highly
non-Gaussian statistics, the bispectrum of such fluctuations should have
nonzero value.  As is well known, PMFs excite not only the scalar
fluctuation but also the vector and tensor fluctuations.  In particular,
it has been known that the vector contribution may dominate over the
scalar one on small scales by the Doppler effect in the CMB power spectrum (e.g. \cite{Mack:2001gc, Lewis:2004ef}).  Hence, the future CMB
experiments, for example, Planck satellite~\cite{:2006uk}, are expected
to give a tighter constraint on the amplitude of the PMFs from the
vector contribution induced from the magnetic fields.  With this
motivation, in Ref. \cite{Shiraishi:2010yk}, we have presented a CMB
angle-averaged bispectrum of the vector perturbations induced from the
PMFs and also a forecast of upper limit for the strength of PMFs
smoothed on $1 {\rm Mpc}$ scale as $B_{\rm 1 Mpc} < 10 {\rm nG}$.
However, there we could not show the details of calculation for the
limit of pages.  Hence, in this paper, we focus on the derivation of the
CMB bispectrum of vector modes induced from PMFs without neglecting the
full angular dependence on the wave number vectors. 
In this paper, we also show that the CMB vector bispectrum induced from the
statistical-isotropic PMFs can be described as the angle-averaged form
like the scalar mode \cite{Komatsu:2001rj} in the rotationally invariant
way 
\footnote{
In Ref.~\cite{Kahniashvili:2010us}, the authors presented the analytical formulas of the CMB vector bispectrum sourced from statistically isotropic PMFs in a different approach than ours and claimed that the bispectrum violates the rotational invariance. Recently, however, they also could reduce the final formulas to the rotational-invariant form, which will be shown in an updated version of their paper~\cite{Kahniashvili:2011pc}.}.

This paper is organized as follows.  In the next section, we formulate
the CMB vector bispectrum induced from PMFs by following the
procedure of Ref. \cite{Shiraishi:2010kd}.  In
Sec. \ref{sec:analytic_calc}, we analytically expand 
the CMB bispectrum with help from some numerical evaluations. In
Sec. \ref{sec:result}, we show our result of the CMB bispectrum from the
PMFs and estimate the limit of the amplitude of the magnetic fields.  In
addition, we also discuss the shape of the bispectrum.  The final
section is devoted to summary and discussion of this paper.

Through this paper, we assume the universe is spatially flat and use the
definition of Fourier transformation:
\begin{eqnarray}
f({\bf x}) \equiv \int \frac{d^3 {\bf k}}{(2 \pi)^3} {f} ({\bf k}) e^{i {\bf k} \cdot {\bf x}}~.
\end{eqnarray}

\section{Formulation of the vector bispectrum induced from PMFs}
\label{sec:formula}

Let us consider the stochastic PMFs $B^b({\bf x}, \tau)$ on the homogeneous
background Universe which is characterized by the
Friedmann-Robertson-Walker metric,
\begin{eqnarray}
ds^2 = a(\tau)^2 \left[ - d\tau^2 + \delta_{bc} dx^b dx^c \right]~,
\end{eqnarray}
where $\tau$ is a conformal time and $a (\tau)$ is a scale factor.  The
expansion of the Universe makes the amplitude of the magnetic fields
decay as $1 / a^2$ and hence we can draw off the time dependence as
$B^b({\bf x}, \tau) = B^b({\bf x}) / a^2$.
The Fourier components of the spatial parts of the PMFs' energy momentum
tensor (EMT) are described as
\begin{eqnarray}
\begin{split}
T^b_{~c}({\bf k},\tau) &\equiv {\rho_\gamma}(\tau)
\left[ \delta^b_{~c} \Delta_B({\bf k}) + \Pi_{Bc}^b({\bf k}) \right]~, \\
\Delta_B({\bf k}) &= {1 \over 8\pi \rho_{\gamma,0}}
\int \frac{d^3 {\bf k}'}{(2\pi)^3} 
B^b({\bf k'}) B_b({\bf k} - {\bf k'})~, \\
\Pi^b_{Bc}({\bf k}) &=-{1 \over 4\pi \rho_{\gamma,0}} \int \frac{d^3 {\bf k'}}{(2 \pi)^3} B^b({\bf k'}) B_c({\bf k} - {\bf k'})~,
\end{split}
\end{eqnarray}
where we have introduced the photon energy density $\rho_{\gamma}$ in
order to include the time dependence of $a^{-4}$ and $\rho_{\gamma, 0}$
denotes the present energy density of photons.  In the following
discussion, for simplicity of calculation, we ignore the trivial
time-dependence. Hence, the index is lowered by
$\delta_{bc}$ and the summation is implied for repeated indices.

Assuming that $B^a({\bf x})$ is a Gaussian field, the
statistically isotropic power spectrum of the PMFs $P_B(k)$ is defined
by
\begin{eqnarray}
\langle B_a({\bf k})B_b({\bf p})\rangle
= (2\pi)^3 {P_B(k) \over 2} P_{ab}(\hat{\bf k}) \delta ({\bf k} + {\bf p})~, \label{eq:def_power}
\end{eqnarray}  
with a projection tensor
\begin{eqnarray}
P_{ab}(\hat{\bf k}) 
\equiv \sum_{\sigma = \pm 1} \epsilon^{(\sigma)}_a(\hat{\bf k}) \epsilon^{(-\sigma)}_b(\hat{\bf k}) 
 =  \delta_{ab} - \hat{k}_a \hat{k}_b~, \label{eq:projection}
\end{eqnarray}
which comes from the divergence free nature of the PMFs.  Here $\hat{\bf
k}$ denotes a unit vector and $\epsilon_a^{(\pm 1)}$ is a normalized
divergenceless polarization vector which satisfies the orthogonal
condition; $\hat{k}^a \epsilon_a^{(\pm 1)} = 0$. The details of the
relations and conventions of the polarization vector are described in
the Appendix in our previous paper~\cite{Shiraishi:2010kd}.  Although the
form of the power spectrum $P_B(k)$ is strongly dependent on the
production mechanism, we assume a simple power law shape given by
\begin{eqnarray}
P_B(k) = A_B k^{n_B}~,
\end{eqnarray}
where $A_B$ and $n_B$ denote the amplitude and the spectral index of the
power spectrum of magnetic fields, respectively.  In order to
parametrize the strength of PMFs, we smooth the magnetic fields with a
conventional Gaussian filter on a comoving scale $r$:
\begin{eqnarray}
B_r^2 \equiv \int_0^\infty \frac{k^2 dk}{2 \pi^2} e^{-k^2 r^2} P_B(k) ~,
\end{eqnarray}
then, $A_B$ is calculated as
\begin{eqnarray}
{A}_B = {
\left(2 \pi \right)^{n_B + 5} B_r^2 \over \Gamma(\frac{n_B + 3}{2}) k_r^{n_B + 3} }~,
\end{eqnarray}
where $\Gamma(x)$ is the Gamma function and $k_r \equiv 2 \pi / r$.

We focus on the vector contribution induced from the PMFs, which comes
from the anisotropic stress of the EMT, i.e., $\Pi_{Bab}$.  Using the
polarization vector, the vector anisotropic stress fluctuation is given
by
\begin{eqnarray}
\Pi_{Bv}^{(\pm 1)}({\bf k}) = \hat{k}_a \epsilon_b^{(\mp 1)}(\hat{\bf k}) \Pi_{Bab}({\bf k})~.
\label{eq:vecani}
\end{eqnarray}
In the magnetic case, this acts as a source of the CMB fluctuations
of vector modes.

\subsection{Bispectrum of the vector anisotropic stress fluctuations}
\label{subsec:bispectrum}

As we have mentioned above, the PMF $B^b$ is assumed
to have Gaussian statistics. Hence one can easily find that the statistics of the
vector anisotropic stress fluctuation given by Eq.~(\ref{eq:vecani}) are
highly non-Gaussian and the bispectrum (3-point function) of that has
a finite value.

Using Eq. (\ref{eq:def_power}) and the Wick's theorem, the bispectrum of
$\Pi_{Bv}^{(\pm 1)}({\bf k})$ is calculated as
\begin{eqnarray}
\Braket{\prod_{n=1}^{3} \Pi_{Bv}^{(\lambda_n)}({\bf k_n})}
&=&   \Braket{\Pi_{Bab} ({\bf k_1}) \Pi_{Bcd} ({\bf k_2})
 \Pi_{Bef} ({\bf k_3})} \hat{k_1}_a \epsilon^{(-\lambda_1)}_b(\hat{\bf k_1})
\hat{k_2}_c \epsilon^{(-\lambda_2)}_d(\hat{\bf k_2})
\hat{k_3}_e \epsilon^{(-\lambda_3)}_f(\hat{\bf k_3}) ~,\label{eq:ini_vec_bis} \\
\Braket{\Pi_{Bab} ({\bf k_1}) \Pi_{Bcd} ({\bf k_2})
 \Pi_{B ef} ({\bf k_3})} 
&=&\left( - 4\pi \rho_{\gamma,0} \right)^{-3}
\left[ \prod_{n=1}^3 \int \frac{d^3 {\bf k_n'}}{(2 \pi)^3} \right]
\nonumber\\
&& \times
\Braket{B_a({\bf k_1'}) B_b({\bf k_1} - {\bf k_1'}) B_c({\bf k_2'}) B_d({\bf k_2}
- {\bf k_2'}) B_e({\bf k_3'}) B_f({\bf k_3} - {\bf k_3'})} \nonumber \\
&=&\left( - 4\pi \rho_{\gamma,0} \right)^{-3} 
\left[ \prod_{n=1}^3 
\int_0^{k_D} k_n'^2 dk_n' P_B(k_n') \int d^2 \hat{\bf k_n'} \right]
\nonumber \\
&& \times 
\delta({\bf k_1} - {\bf k_1'} + {\bf k_3'}) 
\delta({\bf k_2} - {\bf k_2'} + {\bf k_1'}) 
\delta({\bf k_3} - {\bf k_3'} + {\bf k_2'}) \nonumber \\
&& \times \frac{1}{8} [P_{ad}(\hat{\bf k_1'}) P_{be}(\hat{\bf k_3'})
P_{cf}(\hat{\bf k_2'}) + \{a \leftrightarrow b \ {\rm or} \ c \leftrightarrow d \ {\rm or} \ e \leftrightarrow f\}]
~, \label{eq:bis_EMT}
\end{eqnarray}
where $\lambda_n$ denotes the helicity of the vector mode as $\lambda_n
= \pm 1$ and $k_D$ is the Alfv\'en-wave damping length scale
\cite{Jedamzik:1996wp, Subramanian:1997gi} as $k_D^{-1} \sim {\cal
O}(0.1)\rm Mpc$ and the curly bracket denotes the symmetric $7$ terms
under the permutations of indices: $a \leftrightarrow b$, $c
\leftrightarrow d$, or $e \leftrightarrow f$. Note that we express in a
more symmetric form than that of Ref. \cite{Brown:2005kr} to perform the
angular integrals which will be described in
Sec. \ref{sec:analytic_calc}. To avoid the divergence of
$\Braket{\Pi_{Bab} ({\bf k_1}) \Pi_{Bcd} ({\bf k_2}) \Pi_{Bef} ({\bf
k_3})}$ in the IR limit, the value range of the spectral index is
limited as $n_B > -3$.

\subsection{CMB all-sky bispectrum}

The CMB temperature and polarization fluctuations are expanded into
(spin-weighted) spherical harmonics \cite{Zaldarriaga:1996xe,
Lewis:2004ef, Hu:1997hp}.  Then the angle-averaged bispectrum formed by
their coefficient $a^{(Z)}_{X, \ell m}$ can be defined
as~\cite{Komatsu:2001rj, Shiraishi:2010kd}
\begin{eqnarray}
B^{(Z_1 Z_2 Z_3)}_{X_1 X_2 X_3, \ell_1 \ell_2 \ell_3} &\equiv& \sum_{m_1 m_2 m_3}
\left(%
\begin{array}{ccc}
  \ell_1  & \ell_2    & \ell_3   \\
   m_1    & m_2       & m_3   \\
\end{array}%
\right) 
\Braket{\prod_{n=1}^3 a^{(Z_n)}_{X_n, \ell_n m_n}}~,
\label{eq:angleaverage}
\end{eqnarray}
where the matrix is the Wigner-$3j$ symbol, $Z = S, V$ or $T$ is
corresponding to the scalar, vector or tensor-mode perturbation
respectively, and $X = I, E$ or $B$ means intensity, $E$-mode or
$B$-mode polarization, respectively.

Let us consider $B^{(V V V)}_{I I I, \ell_1 \ell_2 \ell_3}$ induced from $\Pi_{Bv}^{(\lambda)}$.
In the same manner as in Refs.~\cite{Shiraishi:2010sm, Shiraishi:2010kd}, 
$a^{(V)}_{I, \ell m}$ sourced from PMF is given by
\begin{eqnarray}
a^{(V)}_{I, \ell m} &=& 4\pi (-i)^{\ell} \int_0^\infty {k^2 dk \over (2\pi)^3} \mathcal{T}^{(V)}_{I, \ell}(k) \sum\limits_{\lambda = \pm 1}
\lambda \Pi_{Bv,\ell
m}^{(\lambda)}(k) ~, \label{eq:alm_general} \\
\Pi_{Bv,\ell m}^{({\pm 1})}(k) &\equiv& \int d^2 \hat{\bf k} \Pi_{Bv}^{({\pm 1})}({\bf k}) 
{}_{\mp 1}Y^*_{\ell m}(\hat{\bf k})~. \label{eq:Pi_expand}
\end{eqnarray}
Here $\mathcal{T}^{(V)}_{I, \ell}$ denotes the radiation transfer
function of the temperature fluctuation from magnetic vector mode as
calculated in Appendix \ref{appen:transfer}, and ${}_{\mp 1}Y_{\ell
m}(\hat{\bf k})$ is the spin-$1$ spherical harmonic function.  By making use
of these equations, we obtain the CMB temperature bispectrum induced
from the vector anisotropic stress $\Pi_{Bv}^{({\pm 1})}$ which is
given by
\begin{eqnarray}
\begin{split}
B^{(V V V)}_{I I I, \ell_1 \ell_2 \ell_3}
&= \left[ \prod\limits^3_{n=1}
4\pi (-i)^{\ell_n}
\int_0^\infty {k_n^2 dk_n \over (2\pi)^3}
\mathcal{T}^{(V)}_{I, \ell_n}(k_n) \right] 
(2\pi)^3 \sum\limits_{\lambda_1, \lambda_2, \lambda_3 = \pm 1}
\lambda_1 \lambda_2 \lambda_3
{\cal F}^{\lambda_1 \lambda_2 \lambda_3}_{\ell_1 \ell_2 \ell_3}(k_1, k_2, k_3)~,  \\
{\cal F}^{\lambda_1 \lambda_2 \lambda_3}_{\ell_1 \ell_2 \ell_3}(k_1, k_2, k_3)
 &\equiv (2 \pi)^{-3}\sum\limits_{m_1 m_2 m_3}
\left(%
\begin{array}{ccc}
  \ell_1  & \ell_2    & \ell_3   \\
   m_1    & m_2       & m_3   \\
\end{array}%
\right) 
\Braket{ \prod_{n=1}^3 \Pi_{Bv,\ell_n m_n}^{(\lambda_n)}(k_n)}
 ~. \label{eq:CMB_vec_bis} 
\end{split}
\end{eqnarray}
These equations are corresponding to Eqs. (2.9) and (2.7) of
Ref. \cite{Shiraishi:2010kd}. 

In the next section, we will derive an explicit form of $B^{(VVV)}_{III,
\ell_1 \ell_2 \ell_3}$ by calculating the complicated angular
dependencies on the wave number vectors, which are implied by
Eqs. (\ref{eq:ini_vec_bis}), (\ref{eq:Pi_expand}) and
(\ref{eq:CMB_vec_bis}), with the spin-weighted spherical harmonics and
the Wigner symbols.  In the calculation, we will see that the dependence
on the azimuthal quantum numbers ($m_1, m_2$ and $m_3$) in the
bispectrum of $\Pi^{(\pm 1)}_{Bv,\ell m}$ is confined only in the same
form as the Wigner-$3j$ symbol in Eq.~(\ref{eq:angleaverage}), which implies
the rotational invariance of the CMB bispectrum from the vector
anisotropic stress of PMFs \cite{Shiraishi:2010yk}.

\section{Analytic calculation of the CMB temperature bispectrum}\label{sec:analytic_calc}
 
In this section, we derive the explicit form of
Eq. (\ref{eq:CMB_vec_bis}) by calculating the full-angular dependence
which has never been considered in the previous studies
\cite{Seshadri:2009sy, Caprini:2009vk, Cai:2010uw, Trivedi:2010gi,
Kahniashvili:2010us}.  The following procedures are based on the
calculation rules discussed in Ref. \cite{Shiraishi:2010kd}. 
Note that we use some colors in the following equations for readers to
follow the complex equations more easily.

\subsection{Exact expression of ${\cal F}^{\lambda_1 \lambda_2 \lambda_3}_{\ell_1 \ell_2 \ell_3}$}

Let us consider an exact expression of ${\cal F}^{\lambda_1 \lambda_2
\lambda_3}_{\ell_1 \ell_2 \ell_3}$, by expanding all the angular
dependencies with spin-weighted spherical harmonics and rewriting the
angular integrals with summations of angular and/or azimuthal quantum
numbers.  Substituting the expression of the bispectrum of
$\Pi_{Bv}^{({\pm 1})}({\bf k})$ (Eq. (\ref{eq:bis_EMT})) and
Eq.~(\ref{eq:Pi_expand}) into Eq.~(\ref{eq:CMB_vec_bis}), we can obtain
\begin{eqnarray}
{\cal F}^{\lambda_1 \lambda_2 \lambda_3}_{\ell_1 \ell_2 \ell_3}(k_1,k_2,k_3) 
&=& 
\sum\limits_{m_1 m_2 m_3}
\left(%
\begin{array}{ccc}
  \ell_1  & \ell_2    & \ell_3   \\
   m_1    & m_2       & m_3   \\
\end{array}%
\right)
\left[ \prod_{n=1}^3 \Purple{ \int d^2 \hat{\bf k_n}
	  {}_{-\lambda_n}Y_{\ell_n m_n}^* (\hat{\bf k_n}) } 
\int^{k_D}_0 k_n'^2 dk_n' P_B(k_n') \Green{\int d^2 \hat{\bf k_n'}}\right] 
 \nonumber \\
&& \times  
\delta({\bf k_1} - {\bf k_1'} + {\bf k_3'}) \delta({\bf k_2} - {\bf k_2'} + {\bf
k_1'}) \delta({\bf k_3} - {\bf k_3'} + {\bf k_2'}) 
\hat{k_1}_a \epsilon^{(- \lambda_1)}_b (\hat{\bf k_1}) 
\hat{k_2}_c \epsilon^{(- \lambda_2)}_d (\hat{\bf k_2})
\hat{k_3}_e \epsilon_f^{(- \lambda_3)} (\hat{\bf k_3})
\nonumber \\
&& \times 
\frac{1}{8} \left[  P_{ad} (\hat{\bf k_1'})  P_{be} (\hat{\bf k_3'}) P_{cf} (\hat{\bf k_2'}) 
 + \{a \leftrightarrow b \ {\rm or} \ c \leftrightarrow d \ {\rm or} \ e
 \leftrightarrow f \}
\right] (-8 \pi^2 \rho_{\gamma,0})^{-3}~.
\label{eq:explicitform}
\end{eqnarray}
At first, we focus on the first term of permutations. 

In the first step,
in order to perform all angular integrals, we expand each function of the wave number vector with the spin-weighted spherical harmonics. 
By this concept, three delta functions are rewritten as
\cite{Shiraishi:2010kd, Shiraishi:2010sm}
\begin{eqnarray}
\begin{split}
\delta({\bf k_1} - {\bf k_1'} + {\bf k_3'}) 
&= 8 \int_0^\infty A^2 dA
\sum _{\substack{L_1 L_2 L_3 \\ \Magenta{M_1 M_2 M_3}}} 
(-1)^{\frac{L_1 + 3 L_2 + L_3}{2}} 
I_{L_1 L_2 L_3}^{0~0~0} 
j_{L_1} (k_1 A) j_{L_2} (k_1' A) j_{L_3} (k_3' A) \\
&\quad \times \Purple{Y_{L_1 M_1}^*(\hat{\bf k_1})} 
\Green{ Y_{L_2 M_2}(\hat{\bf k_1'}) Y_{L_3 M_3}^*(\hat{\bf k_3'}) }
\Magenta{ (-1)^{M_2} \left(
  \begin{array}{ccc}
  L_1 &  L_2 & L_3 \\
   M_1 & -M_2 & M_3
  \end{array}
 \right) }~, \\
\delta({\bf k_2} - {\bf k_2'} + {\bf k_1'}) 
&= 8 \int_0^\infty B^2 dB 
\sum _{\substack{L_1' L_2' L_3' \\ \Orange{M_1' M_2' M_3'}}} 
(-1)^{\frac{L_1' + 3 L_2' + L_3'}{2}}  
I_{L'_1 L'_2 L'_3}^{0~0~0} 
j_{L'_1} (k_2 B) j_{L'_2} (k_2' B) j_{L'_3} (k_1'B) \\
&\quad \times \Purple{Y_{L'_1 M'_1}^*(\hat{\bf k_2}) }
\Green{ Y_{L'_2 M'_2}(\hat{\bf k_2'}) Y_{L'_3 M'_3}^*(\hat{\bf k_1'}) }
\Orange{ (-1)^{M'_2} \left(
  \begin{array}{ccc}
  L'_1 &  L'_2 & L'_3 \\
   M'_1 & -M'_2 & M'_3
  \end{array}
 \right) }~, \\
\delta({\bf k_3} - {\bf k_3'} + {\bf k_2'}) 
&= 8 \int_0^\infty C^2 dC 
\sum _{\substack{L_1'' L_2'' L_3'' \\ \Brown{M_1'' M_2'' M_3''}}} 
(-1)^{\frac{L''_1 + 3 L''_2 + L''_3}{2}} 
I_{L''_1 L''_2 L''_3}^{0~0~0} 
j_{L''_1} (k_3 C) j_{L''_2} (k_3' C) j_{L''_3} (k_2 'C) \\
&\quad \times \Purple{Y_{L''_1 M''_1}^*(\hat{\bf k_3}) }
\Green{ Y_{L''_2 M''_2}(\hat{\bf k_3'}) Y_{L''_3 M''_3}^*(\hat{\bf k_2'}) }
\Brown{ (-1)^{M''_2} \left(
  \begin{array}{ccc}
  L''_1 &  L''_2 & L''_3 \\
   M''_1 & -M''_2 & M''_3
  \end{array}
 \right) }~, \label{eq:delta_C}
\end{split}
\end{eqnarray}
where
\begin{eqnarray}
I^{s_1 s_2 s_3}_{\ell_1 \ell_2 \ell_3} 
\equiv \sqrt{\frac{(2 \ell_1 + 1)(2 \ell_2 + 1)(2 \ell_3 + 1)}{4 \pi}}
\left(
  \begin{array}{ccc}
  \ell_1 & \ell_2 & \ell_3 \\
   s_1 & s_2 & s_3 
  \end{array}
 \right)~.
\end{eqnarray}
The other functions in Eq.~(\ref{eq:explicitform}), which depend on the
angles of the wave number ${\bf k}_n$, can be also expanded in terms
of the spin-weighted spherical harmonics as
 \begin{eqnarray}
\begin{split}
\hat{k_1}_a \epsilon^{(- \lambda_2)}_d (\hat{\bf k_2}) P_{ad} (\hat{\bf k_1'}) 
&= \hat{k_1}_a \epsilon^{(- \lambda_2)}_d (\hat{\bf k_2}) 
\sum_{\sigma = \pm 1} \epsilon^{(\sigma)}_a (\hat{\bf k_1'}) \epsilon^{(-
 \sigma)}_d (\hat{\bf k_1'}) \\
&= \sum_{\sigma = \pm 1} \sum_{\Magenta{m_a}, \Orange{m_d}} \left(\frac{4 \pi}{3}\right)^2 
(- \lambda_2) \Purple{ Y_{1 m_a} (\hat{\bf k_1}) {}_{- \lambda_2}Y_{1 m_d}
 (\hat{\bf k_2}) }
\Green{ {}_{- \sigma}Y^*_{1 m_a} (\hat{\bf k_1'}) {}_{\sigma}Y^*_{1 m_d}
 (\hat{\bf k_1'}) }~, \\
\hat{k_2}_c  \epsilon_f^{(- \lambda_3) }(\hat{\bf k_3}) P_{cf} (\hat{\bf k_2'})
&=  \sum_{\sigma'' = \pm 1} \sum_{\Orange{m_c}, \Brown{m_f}} \left(\frac{4 \pi}{3}\right)^2 
(- \lambda_3) \Purple{ Y_{1 m_c} (\hat{\bf k_2}) {}_{- \lambda_3}Y_{1 m_f}
 (\hat{\bf k_3}) } 
\Green{ {}_{- \sigma''}Y^*_{1 m_c} (\hat{\bf k_2'}) {}_{\sigma''}Y^*_{1
 m_f} (\hat{\bf k_2'}) }~,
\\
\hat{k_3}_e  \epsilon^{(- \lambda_1)}_b (\hat{\bf k_1})  P_{be} (\hat{\bf k_3'})  
&= \sum_{\sigma' = \pm 1} \sum_{\Magenta{m_b}, \Brown{m_e}} \left(\frac{4 \pi}{3}\right)^2 
(- \lambda_1) \Purple{ Y_{1 m_e} (\hat{\bf k_3}) {}_{- \lambda_1}Y_{1 m_b}
 (\hat{\bf k_1}) }
\Green{ {}_{- \sigma'}Y^*_{1 m_e} (\hat{\bf k_3'}) {}_{\sigma'}Y^*_{1 m_b}
 (\hat{\bf k_3'}) }~,
\end{split}
\end{eqnarray}
where we have used Eq. (\ref{eq:projection}) and the notations of a unit
vector $\hat{\bf n}$ and a divergenceless unit vector $\epsilon_a^{(\pm
1)}$ as \cite{Shiraishi:2010kd}
\begin{eqnarray}
\begin{split}
\hat{n}_a &= \sum_m \alpha_a^{m} Y_{1 m}(\hat{\bf n})~, \\
\epsilon_a^{(\pm 1)} (\hat{\bf n}) 
&= \epsilon_a^{(\mp 1) *} (\hat{\bf n}) 
= \mp \sum_m \alpha_a^m {}_{\pm 1} Y_{1 m} (\hat{\bf n})~, \\
\alpha_a^m \alpha_a^{m'} &= \frac{4 \pi}{3} (-1)^m \delta_{m, -m'} ~.
\end{split}
\end{eqnarray} 

In the second step, let us consider performing all angular integrals and
replacing them with the Wigner-$3j$ symbols.  Three angular integrals with
respect to $\hat{\bf k_1'}, \hat{\bf k_2'}$ and $\hat{\bf k_3'}$ are
given as
\begin{eqnarray}
\begin{split}
 \Green{
\int d^2 \hat{\bf k_1'} {}_{-\sigma} Y^*_{1 m_a} Y_{L_2 M_2} 
{}_\sigma Y^*_{1 m_d} Y^*_{L'_3 M'_3}} 
&= \sum_{L \Cyan{M}} \sum_{S = \pm 1} (-1)^{\sigma + \Magenta{m_a}} I_{L_3' 1 L}^{0 -\sigma -S} I_{L_2 1 L}^{0 -\sigma -S} 
\Orange{ \left(
  \begin{array}{ccc}
  L'_3 &  1 & L \\
   M'_3 & m_d & M 
  \end{array}
 \right) } 
\Magenta{ \left(
  \begin{array}{ccc}
  L_2 &  1 & L \\
   M_2 & -m_a & M 
  \end{array}
 \right) }~, \\
\Green{
\int d^2 \hat{\bf k_2'} {}_{- \sigma''} Y^*_{1 m_c} Y_{L'_2 M'_2} 
{}_{\sigma''} Y^*_{1 m_f} Y^*_{L''_3 M''_3} } 
&= \sum_{L' \Cyan{M'}} \sum_{S' = \pm 1} (-1)^{\sigma'' + \Orange{m_c}} I_{L_3'' 1 L'}^{0 -\sigma'' -S'} I_{L'_2 1 L'}^{0 -\sigma'' -S'} 
\Brown{ \left(
  \begin{array}{ccc}
  L''_3 &  1 & L' \\
   M''_3 & m_f & M' 
  \end{array}
 \right) } 
\Orange{ \left(
  \begin{array}{ccc}
  L'_2 &  1 & L' \\
   M'_2 & -m_c & M' 
  \end{array}
 \right) }~, \\
\Green{
\int d^2 \hat{\bf k_3'} 
{}_{- \sigma'} Y^*_{1 m_e} Y_{L''_2 M''_2} 
{}_{\sigma'} Y^*_{1 m_b} Y^*_{L_3 M_3}
} 
&= \sum_{L'' \Cyan{M''}} \sum_{S'' = \pm 1} (-1)^{\sigma' + \Brown{m_e}} I_{L_3 1 L''}^{0 -\sigma' -S''} I_{L''_2 1 L''}^{0 -\sigma' -S''} 
\Magenta{ \left(
  \begin{array}{ccc}
  L_3 &  1 & L'' \\
   M_3 & m_b & M'' 
  \end{array}
 \right)  }
\Brown{ \left(
  \begin{array}{ccc}
  L''_2 &  1 & L'' \\
   M''_2 & -m_e & M'' 
  \end{array}
 \right) }~, \label{eq:ang_int_kdpdqd}
\end{split}
\end{eqnarray}
where we have used a property of spin-weighted spherical harmonics given
by~\cite{Hu:2001fa, Shiraishi:2010kd}
\begin{eqnarray}
\prod_{n=1}^2 {}_{s_n} Y_{l_n m_n}
 &=& \sum_{l_3 m_3 s_3} {}_{s_3} Y^*_{l_3 m_3} 
I^{-s_1 -s_2
-s_3}_{l_1~l_2~l_3} 
\left(
  \begin{array}{ccc}
  l_1 & l_2 & l_3 \\
  m_1 & m_2 & m_3
  \end{array}
 \right)~, \label{eq:product_sYlm} \\
{}_sY_{l m}^*  &=& (-1)^{s+m} {}_{-s}Y_{l -m} ~.
\end{eqnarray}
We can also perform the angular integrals with respect to $\hat{\bf
k_1}, \hat{\bf k_2}$ and $\hat{\bf k_3}$ as
\begin{eqnarray}
\begin{split}
\Purple{
\int d^2 \hat{\bf k_1} 
{}_{-\lambda_1} Y_{1 m_b} Y_{1 m_a}
{}_{-\lambda_1} Y^*_{\ell_1 m_1} Y^*_{L_1 M_1}
} 
&= \sum_{L_k \Magenta{M_k}}\sum_{S_k = \pm 1} I_{L_1 \ell_1 L_k}^{0 \lambda_1 -S_k} I_{1 1 L_k}^{0 \lambda_1 -S_k} 
\Magenta{ \left(
  \begin{array}{ccc}
  L_1 & \ell_1 & L_k \\
  M_1 & m_1 & M_k 
  \end{array}
 \right)  
\left(
  \begin{array}{ccc}
  1 &  1 & L_k \\
  m_a & m_b & M_k 
  \end{array}
 \right) } ~,\\
\Purple{
\int d^2 \hat{\bf k_2} {}_{-\lambda_2} Y_{1 m_d} Y_{1 m_c} 
{}_{-\lambda_2} Y^*_{\ell_2 m_2} Y^*_{L'_1 M'_1}
} 
&= \sum_{L_p \Orange{M_p}}\sum_{S_p = \pm 1} I_{L'_1 \ell_2 L_p}^{0 \lambda_2 -S_p} I_{1 1 L_p}^{0 \lambda_2 -S_p} 
\Orange{ \left(
  \begin{array}{ccc}
  L'_1 & \ell_2 & L_p \\
  M'_1 & m_2 & M_p 
  \end{array}
 \right)  
\left(
  \begin{array}{ccc}
  1 &  1 & L_p \\
  m_c & m_d & M_p 
  \end{array}
 \right) } ~,\\
\Purple{
\int d^2 \hat{\bf k_3} 
{}_{- \lambda_3} Y_{1 m_f}  Y_{1 m_e}
{}_{- \lambda_3} Y^*_{\ell_3 m_3} Y^*_{L''_1 M''_1}
}
&= \sum_{L_q \Brown{M_q}}\sum_{S_q = \pm 1} I_{L''_1 \ell_3 L_q}^{0 \lambda_3 -S_q} I_{1 1 L_q}^{0 \lambda_3 -S_q}
\Brown{ \left(
  \begin{array}{ccc}
  L''_1 & \ell_3 & L_q \\
  M''_1 & m_3 & M_q 
  \end{array}
 \right)  
\left(
  \begin{array}{ccc}
  1 &  1 & L_q \\
  m_e & m_f & M_q 
  \end{array}
 \right) }~. \label{eq:ang_int_kpq}
\end{split} 
\end{eqnarray} 
At this point, all the angular integrals in Eq.~(\ref{eq:explicitform})
have been reduced into the Wigner-$3j$ symbols.

Then, in the third step, we consider summing up the Wigner-$3j$ symbols
in terms of the azimuthal quantum numbers and replacing them with the
Wigner-$6j$ and $9j$ symbols, which denote Clebsch-Gordan coefficients
between two other eigenstates coupled to three and four individual
momenta \cite{Hu:2001fa, Gurau:2008, Jahn/Hope:1954, Shiraishi:2010kd}.
Using these properties, we can express the summation of five Wigner-$3j$
symbols with a Wigner-$9j$ symbol:
\begin{eqnarray}
\begin{split}
&\Magenta{
\sum_{\substack{M_1 M_2 M_3 \\ M_k m_a m_b}} (-1)^{M_2+m_a} 
\left(
  \begin{array}{ccc}
  L_1 & L_2 & L_3 \\
  M_1 & - M_2 & M_3 
  \end{array}
 \right)
\left(
  \begin{array}{ccc}
  1 & 1 & L_k \\
  m_a & m_b & M_k 
  \end{array}
 \right)
\left(
  \begin{array}{ccc}
  L_3 & 1 & L'' \\
  M_3 & m_b & M'' 
  \end{array}
 \right)
 \left(
  \begin{array}{ccc}
  L_2 &  1 & L \\
  M_2 & -m_a & M 
  \end{array}
 \right)
\left(
  \begin{array}{ccc}
  L_1 & \ell_1 & L_k \\
  M_1 & m_1 & M_k 
  \end{array}
 \right)
} \\
&\qquad\qquad = (-1)^{\Cyan{M} + \ell_1 + L_3 + L + 1} 
\Cyan{ \left(
  \begin{array}{ccc}
  L'' & L & \ell_1 \\
  M'' & -M & m_1 
  \end{array}
 \right) }
\left\{
  \begin{array}{ccc}
  L'' & L & \ell_1 \\
  L_3 & L_2 & L_1 \\
  1 & 1 & L_k 
  \end{array}
 \right\}~, \\
&\Orange{
\sum_{\substack{M'_1 M'_2 M'_3 \\ M_p m_c m_d} } (-1)^{M'_2 + m_c} 
\left(
  \begin{array}{ccc}
  L'_1 & L'_2 & L'_3 \\
  M'_1 & - M'_2 & M'_3 
  \end{array}
 \right)
\left(
  \begin{array}{ccc}
  1 & 1 & L_p \\
  m_c & m_d & M_p 
  \end{array}
 \right) 
\left(
  \begin{array}{ccc}
  L'_3 & 1 & L \\
  M'_3 & m_d & M 
  \end{array}
 \right)
 \left(
  \begin{array}{ccc}
  L'_2 &  1 & L' \\
  M'_2 & -m_c & M' 
  \end{array}
 \right)
\left(
  \begin{array}{ccc}
  L'_1 & \ell_2 & L_p \\
  M'_1 & m_2 & M_p 
  \end{array}
 \right)
} \\
&\qquad\qquad= (-1)^{\Cyan{M'} + \ell_2 + L'_3 + L' + 1} 
\Cyan{ \left(
  \begin{array}{ccc}
  L & L' & \ell_2 \\
  M & -M' & m_2 
  \end{array}
 \right) }
\left\{
  \begin{array}{ccc}
  L & L' & \ell_2 \\
  L'_3 & L'_2 & L'_1 \\
  1 & 1 & L_p 
  \end{array}
 \right\}~, \\
&\Brown{
\sum_{\substack{M''_1 M''_2 M''_3 \\ M_q m_e m_f}} (-1)^{M''_2 + m_e} 
\left(
  \begin{array}{ccc}
  L''_1 & L''_2 & L''_3 \\
  M''_1 & - M''_2 & M''_3 
  \end{array}
 \right)
\left(
  \begin{array}{ccc}
  1 & 1 & L_q\\
  m_e & m_f & M_q 
  \end{array}
 \right) 
\left(
  \begin{array}{ccc}
  L''_3 & 1 & L' \\
  M''_3 & m_f & M' 
  \end{array}
 \right)
 \left(
  \begin{array}{ccc}
  L''_2 &  1 & L'' \\
  M''_2 & -m_e & M'' 
  \end{array}
 \right)
\left(
  \begin{array}{ccc}
  L''_1 & \ell_3 & L_q \\
  M''_1 & m_3 & M_q 
  \end{array}
 \right)
} \\
&\qquad\qquad= (-1)^{\Cyan{M''} + \ell_3 + L''_3 + L'' + 1} 
\Cyan{ \left(
  \begin{array}{ccc}
  L' & L'' & \ell_3 \\
  M' & -M'' & m_3 
  \end{array}
 \right) }
\left\{
  \begin{array}{ccc}
  L' & L'' & \ell_3 \\
  L''_3 & L''_2 & L''_1 \\
  1 & 1 & L_q 
  \end{array}
 \right\} ~.
\end{split} \label{eq:sum_ms}
\end{eqnarray}
Furthermore, we can also sum up the renewed Wigner-$3j$ symbols 
arising in the above equations
over $M, M'$ and $M''$ with the Wigner-$6j$ symbol as \cite{mathematica}
\begin{eqnarray}
&& \Cyan{
\sum_{M M' M''} (-1)^{M + M' + M''} 
\left(
  \begin{array}{ccc}
  L'' & L & \ell_1 \\
  M'' & -M & m_1 
  \end{array}
 \right)
\left(
  \begin{array}{ccc}
  L & L' & \ell_2 \\
  M & -M' & m_2 
  \end{array}
 \right)
\left(
  \begin{array}{ccc}
  L' & L'' & \ell_3 \\
  M' & -M'' & m_3 
  \end{array}
 \right)} \nonumber \\
&&\qquad\qquad = (-1)^{L+L'+L''}
\left(
  \begin{array}{ccc}
  \ell_1 & \ell_2 & \ell_3 \\
  m_1 & m_2 & m_3 
  \end{array}
 \right)
\left\{
  \begin{array}{ccc}
  \ell_1 & \ell_2 & \ell_3 \\
  L' & L'' & L 
  \end{array}
 \right\}~.
\end{eqnarray}   
With this prescription, one can find that the three azimuthal numbers
are confined only in the Wigner-$3j$ symbol as $\left(
  \begin{array}{ccc}
  \ell_1 & \ell_2 & \ell_3 \\
  m_1 & m_2 & m_3 
  \end{array}
 \right)$.
 This $3j$ symbol arises from the bispectrum of $\Pi^{(\pm
 1)}_{Bv, \ell m}$
and exactly ensures the rotational invariance of the CMB
 bispectrum as pointed out above.

So far, we have considered only the first term of permutations in
Eq.~(\ref{eq:explicitform}).  Hence, finally, we have to consider the
contribution of the other $7$ permutations. For example, in the
calculation of the $\{a \leftrightarrow b\}$ part, $m_a$ and $m_b$ of
Eq. (\ref{eq:ang_int_kdpdqd}) replace each other and the summation over
$m_a$ and $m_b$ in Eq. (\ref{eq:sum_ms}) changes as
\begin{eqnarray}
&& \sum_{\substack{M_1 M_2 M_3 \\ M_k m_b m_a}} (-1)^{M_2+m_b} 
\left(
  \begin{array}{ccc}
  L_1 & L_2 & L_3 \\
  M_1 & - M_2 & M_3 
  \end{array}
 \right)
\left(
  \begin{array}{ccc}
  1 & 1 & L_k \\
  m_a & m_b & M_k 
  \end{array}
 \right)
\left(
  \begin{array}{ccc}
  L_3 & 1 & L'' \\
  M_3 & m_a & M'' 
  \end{array}
 \right)
 \left(
  \begin{array}{ccc}
  L_2 &  1 & L \\
  M_2 & -m_b & M 
  \end{array}
 \right)
\left(
  \begin{array}{ccc}
  L_1 & \ell_1 & L_k \\
  M_1 & m_1 & M_k 
  \end{array}
 \right) \nonumber \\
&& \qquad\qquad = (-1)^{M + \ell_1 + L_3 + L + 1 + L_k} 
\left(
  \begin{array}{ccc}
  L'' & L & \ell_1 \\
  M'' & -M & m_1 
  \end{array}
 \right) 
\left\{
  \begin{array}{ccc}
  L'' & L & \ell_1 \\
  L_3 & L_2 & L_1 \\
  1 & 1 & L_k 
  \end{array}
 \right\}~,
\end{eqnarray}
hence, the extra factor $(-1)^{L_k}$ arises.  In the same manner, we can
find the extra factor $(-1)^{L_p}$ or $(-1)^{L_q}$ in the $\{ c
\leftrightarrow d \}$ or the $\{e \leftrightarrow f \}$ part,
respectively.

Using the above expansions and the orthogonality of the Wigner-$3j$
symbols given by
\begin{eqnarray}
\displaystyle \sum_{m_1 m_2 m_3} \left(
  \begin{array}{ccc}
  \ell_1 & \ell_2 & \ell_3 \\
  m_1 & m_2 & m_3 
  \end{array}
 \right)^2 = 1~,
\end{eqnarray} 
and performing the summations over $L_k, L_p$ and $L_q$ such as
\begin{eqnarray}
\sum_{L_k} I_{L_1 \ell L_k}^{0 \lambda -\lambda}
 I_{1 1 L_k}^{0 \lambda -\lambda} 
\frac{1 + (-1)^{L_k}}{2}
 \left\{
  \begin{array}{ccc}
  L'' & L & \ell \\
  L_3 & L_2 & L_1 \\
  1 & 1 & L_k
  \end{array}
 \right\} 
= - \frac{3}{2 \sqrt{2 \pi}} 
I_{L_1 \ell 2}^{0 \lambda -\lambda}
 \left\{
  \begin{array}{ccc}
  L'' & L & \ell \\
  L_3 & L_2 & L_1 \\
  1 & 1 & 2
  \end{array}
 \right\} ~, 
\end{eqnarray}
we can obtain an exact form of ${\cal F}^{\lambda_1 \lambda_2
\lambda_3}_{\ell_1 \ell_2 \ell_3}$ given by
\begin{eqnarray}
{\cal F}^{\lambda_1 \lambda_2 \lambda_3}_{\ell_1 \ell_2 \ell_3}(k_1,k_2,k_3)
&=& (-8 \pi^2 \rho_{\gamma,0})^{-3}\left[ \prod_{n=1}^3 \int_0^{k_D} k_n'^2 dk_n' P_B(k_n') \right]
\nonumber \\ 
&& \times \sum_{L L' L''} \sum_{S, S', S'' = \pm 1} 
\left\{
  \begin{array}{ccc}
  \ell_1 & \ell_2 & \ell_3 \\
  L' & L'' & L 
  \end{array}
 \right\}
f^{S'' S \lambda_1}_{L'' L \ell_1 }(k_3',k_1',k_1) f^{S S' \lambda_2}_{L L' \ell_2}(k_1',k_2',k_2)
f^{S' S'' \lambda_3}_{L' L'' \ell_3}(k_2',k_3',k_3),
 \label{eq:calF_exact}
\end{eqnarray}
where 
\begin{eqnarray}
f^{S'' S \lambda}_{L'' L \ell}(r_3, r_2, r_1) 
&=& \frac{2(8\pi)^{3/2}}{3}
\sum_{L_1 L_2 L_3} \int_0^\infty A^2 dA j_{L_3}(r_3 A) j_{L_2}(r_2 A) j_{L_1}(r_1 A)  \nonumber \\
&& \times
 \lambda (-1)^{\ell + L_2+L_3} (-1)^{\frac{L_1 + L_2 + L_3}{2}} 
I^{0~0~0}_{L_1 L_2 L_3} I^{0 S'' -S''}_{L_3 1 L''} I^{0 S -S}_{L_2 1 L} 
I_{L_1 \ell 2}^{0 \lambda -\lambda} 
 \left\{
  \begin{array}{ccc}
  L'' & L & \ell \\
  L_3 & L_2 & L_1 \\
  1 & 1 & 2
  \end{array}
 \right\}~. \label{eq:f_exact}
\end{eqnarray}
This expression is one of our results in this paper.  The above
analytic expression of ${\cal F}^{\lambda_1 \lambda_2 \lambda_3}_{\ell_1
\ell_2 \ell_3}$ seems to be quite useful to calculate the CMB bispectrum
of vector modes induced from PMFs with the full angular dependence.
However, it is still too hard to calculate numerically,  because the full
expression of the bispectrum has six integrals and summations over the
helicities as 
\begin{eqnarray}
B^{(V V V)}_{I I I, \ell_1 \ell_2 \ell_3}
&=& 
\left( -\frac{8 (2\pi)^{1/2}}{ 3 \rho_{\gamma,0}} \right)^3 
\sum _{L L' L''} 
\left\{
  \begin{array}{ccc}
  \ell_1 & \ell_2 & \ell_3 \\
  L' & L'' & L 
  \end{array}
 \right\}
\nonumber \\
&& \times
\sum_{\substack{L_1 L_2 L_3 \\ L'_1 L'_2 L'_3 \\ L''_1 L''_2 L''_3}} 
(-1)^{\sum_{i=1}^3\frac{L_i+L'_i+L''_i+2 \ell_i}{2}} 
I^{0~0~0}_{L_1 L_2 L_3} I^{0~0~0}_{L'_1 L'_2 L'_3} I^{0~0~0}_{L''_1
L''_2 L''_3}  
\left\{
  \begin{array}{ccc}
  L'' & L & \ell_1 \\
  L_3 & L_2 & L_1 \\
  1 & 1 & 2
  \end{array}
 \right\}
\left\{
  \begin{array}{ccc}
  L & L' & \ell_2 \\
  L'_3 & L'_2 & L'_1 \\
  1 & 1 & 2
  \end{array}
 \right\}
\left\{
  \begin{array}{ccc}
  L' & L'' & \ell_3 \\
  L''_3 & L''_2 & L''_1 \\
  1 & 1 & 2
  \end{array}
 \right\} \nonumber \\
&& \times 
\Blue{
\left[ \prod\limits^3_{i=1}
4\pi (-i)^{\ell_i}
\int_0^\infty {k_i^2 dk_i \over (2\pi)^3}
\mathcal{T}^{(V)}_{I, \ell_i}(k_i)\right] 
\int_0^\infty A^2 dA j_{L_1} (k_1 A)  
\int_0^\infty B^2 dB j_{L'_1} (k_2 B)  
\int_0^\infty C^2 dC j_{L''_1} (k_3 C) \nonumber \\ 
&& \times \int_0^{k_D} k_1'^2 dk_1' P_B(k_1') 
j_{L_2} (k_1' A) j_{L'_3} (k_1' B) 
 \int_0^{k_D} k_2'^2 dk_2' P_B(k_2')
j_{L'_2} (k_2'B) j_{L''_3} (k_2'C) \nonumber \\
&& \times 
\int_0^{k_D} k_3'^2 dk_3'  P_B(k_3') 
j_{L''_2} (k_3'C) j_{L_3} (k_3'A) }\nonumber \\
&& \times 
\Red{\sum_{S, S', S'' = \pm 1} (-1)^{L_2 + L'_2 + L''_2 + L_3 + L'_3 + L''_3}
I^{0 S -S}_{L'_3 1 L} I^{0 S -S}_{L_2 1 L} I^{0 S' -S'}_{L''_3 1 L'} 
I^{0 S' -S'}_{L'_2 1 L'} I^{0 S'' -S''}_{L_3 1 L''} I^{0 S'' -S''}_{L''_2 1 L''}} \nonumber \\
&& \times 
\OliveGreen{ \sum_{\lambda_1, \lambda_2, \lambda_3 = \pm 1} 
I^{0 \lambda_1 -\lambda_1}_{L_1 \ell_1 2} 
I^{0 \lambda_2 -\lambda_2}_{L'_1 \ell_2 2} 
I^{0 \lambda_3 -\lambda_3}_{L''_1 \ell_3 2} }~.
\label{eq:fullbispectrum}
\end{eqnarray}
In the following subsection, we introduce an approximation, the so-called,
thin last scattering surface (LSS) approximation to reduce the integrals
and perform the summations over the helicities based on the
selection rules for Wigner-$3$j symbols.

\subsection{Thin LSS approximation}

\begin{figure}[t]
  \begin{center}
    \includegraphics{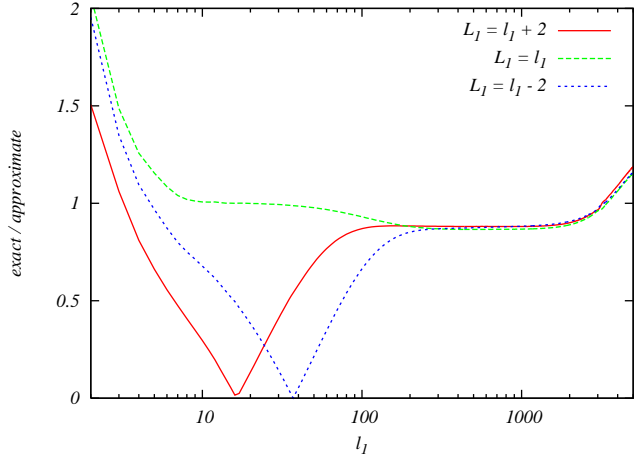}
  \end{center}
  \caption{(color online). The ratio of the left-hand side (exact solution) to the right-hand side (approximate solution) in Eq. (\ref{eq:thin_LSS_check}). The lines correspond to the case for $L_1 = \ell_1 + 2$ (red solid line), for $L_1 = \ell_1$ (green dashed line), and for $L_1 = \ell_1 - 2$ (blue dotted line).}
  \label{fig:thin_LSS_check.eps}
\end{figure}

Let us consider the parts of the integrals with respect to $A, B, C, k',
p'$ and $q'$ in the full expression of the
bispectrum~(Eq.(\ref{eq:fullbispectrum})).
In the computation of the CMB bispectrum, the integral in terms of $k$
(, $p$ and $q$) appears in the form as $\int k^2 dk {\cal
T}^{(V)}_{I,\ell_1}(k) j_{L_1}(k A)$.  We find that this integral is
sharply-peaked at $A \simeq \tau_0 -\tau_*$, where $\tau_0$ is the
present conformal time and $\tau_*$ is the conformal time of the
recombination epoch. According to Ref. \cite{Mack:2001gc, Lewis:2004ef},
the vorticity of subhorizon scale sourced by magnetic fields around the
recombination epoch mostly contributes to generate the CMB vector
perturbation. On the other hand, since the vector mode in the metric
decays after neutrino decoupling, the integrated Sachs-Wolfe effect
after recombination is not observable.  Such a behavior of the
transfer function would be understood based on the calculation in
Appendix \ref{appen:transfer} and we expect ${\cal T}^{(V)}_{I,\ell_1}(k)
\propto j_{\ell_1}(k(\tau_0 - \tau_*))$,  and the $k$-integral behaves
like $\delta(A - (\tau_0 - \tau_*))$.  By the numerical computation, we
found that
\begin{eqnarray}
\int^{\infty}_0 A^2 dA \int_0^\infty k_1^2 dk_1 {\cal T}^{(V)}_{I,\ell_1}(k_1) j_{L_1}(k_1 A)
\simeq  (\tau_0 - \tau_*)^2 \left({\tau_* \over 5}\right) \int k_1^2 dk_1 {\cal T}^{(V)}_{I,\ell_1}(k_1) j_{\ell_1}(k_1 (\tau_0 - \tau_*))~, \label{eq:thin_LSS_check}
\end{eqnarray} 
is a good approximation for $L_1 = \ell_1 \pm 2, \ell_1$ as described in
Fig. \ref{fig:thin_LSS_check.eps}. Note that only the cases $L_1 =
\ell_1 \pm 2, \ell_1$ should be considered due to the selection rules
for Wigner-3j symbols as we shall see later. From this figure, we can
find that the approximation (the right-handed term of
Eq. (\ref{eq:thin_LSS_check})) has less than $20 \%$ uncertainty for
$\ell_1 \simeq L_1 \gtrsim 100$, and therefore this approximation leads
to only less than $10 \%$ uncertainty in the bound on the strength of PMFs
if we place the constraint from the bispectrum data at $\ell_1,
\ell_2 ,\ell_3 \gtrsim 100$ \footnote{ Of course, if we calculate the
bispectrum at smaller multipoles, we may perform the full integration
without this approximation }.
Using this approximation, namely $A = B = C \rightarrow \tau_0 - \tau_*$
and $\int dA = \int dB = \int dC \rightarrow \tau_*/5$, the integrals
with respect to $A,B,C,k', p'$ and $q'$ are estimated as
\begin{eqnarray}
&&\Blue{
\left[ \prod\limits^3_{n=1}
4\pi (-i)^{\ell_n}
\int_0^\infty {k_n^2 dk_n \over (2\pi)^3}
\mathcal{T}^{(V)}_{I, \ell_n}(k_n)\right] 
\int_0^\infty A^2 dA j_{L_1} (k_1 A)  
\int_0^\infty B^2 dB j_{L'_1} (k_2 B)  
\int_0^\infty C^2 dC j_{L''_1} (k_3 C) \nonumber \\ 
&& \qquad\qquad \times 
 \int_0^{k_D} k_1'^2 dk_1'  P_B(k_1') 
j_{L_2} (k_1'A) j_{L'_3} (k_1' B) 
 \int_0^{k_D} k_2'^2 dk_2' P_B(k_2')
j_{L'_2} (k_2'B) j_{L''_3} (k_2'C) \nonumber \\
&& \qquad\qquad\times 
\int_0^{k_D} k_3'^2 dk_3'  P_B(k_3') 
j_{L''_2} (k_3'C) j_{L_3} (k_3'A) } \nonumber \\
&& \qquad \simeq
\left[ \prod\limits^3_{n=1}
4\pi (-i)^{\ell_n}
\int_0^\infty {k_n^2 dk_n \over (2\pi)^3}
\mathcal{T}^{(V)}_{I, \ell_n}(k_n)
j_{\ell_n} (k_n (\tau_0-\tau_*))
\right] \nonumber \\
&& \qquad\qquad\qquad \times 
 A_B^3 (\tau_0-\tau_*)^6 \left( \frac{\tau_*}{5} \right)^3 
\mathcal{K}_{L_2 L'_3}^{-(n_B +1)}(\tau_0-\tau_*)
\mathcal{K}_{L'_2 L''_3}^{-(n_B +1)}(\tau_0-\tau_*)
\mathcal{K}_{L''_2 L_3}^{-(n_B +1)}(\tau_0-\tau_*) ~.
\end{eqnarray}
Here the function ${\cal K}^N_{ll'}$ is defined as 
\begin{eqnarray}
\mathcal{K}_{l l'}^{N}(y) 
&\equiv& \int_0^\infty dk k^{1 - N} j_{l} (ky) j_{l'} (ky) \nonumber \\
&=& \frac{\pi}{2y} \frac{y^{N-1}}{2^N} \frac{\Gamma(N) \Gamma(\frac{l + l' + 2 - N}{2})}{\Gamma(\frac{l - l' + 1 + N}{2}) \Gamma(\frac{-l + l' + 1 + N}{2}) 
\Gamma(\frac{l + l' + 2 + N}{2})} \ ({\rm for} \ y,N, l+l'+2-N>0)~,
\end{eqnarray} 
which behaves asymptotically as ${\cal K}_{l l'}^{N}(y) \propto l^{-N}$
for $l \sim l' \gg 1$.  Here we have evaluated the $k^\prime$ integrals
by setting $k_D \rightarrow \infty$.
This is also a good approximation because the integrands are suppressed
enough for $k', p', q' < k_D \sim {\cal O}(10) \rm Mpc^{-1}$.

\subsection{Selection rules of the Wigner-$3j$ symbol}

Next we consider performing the summations with respect to the
helicities of vector modes.  By considering the selection rules of the
Wigner-$3j$ symbol, the summations over $S$, $S'$ and $S''$ (red part in
Eq.~(\ref{eq:fullbispectrum})) are performed as
\begin{eqnarray}
&&\Red{\sum_{S, S', S'' = \pm 1} (-1)^{L_2 + L'_2 + L''_2 + L_3 + L'_3 + L''_3}
I^{0 S -S}_{L'_3 1 L} I^{0 S -S}_{L_2 1 L} I^{0 S' -S'}_{L''_3 1 L'} 
I^{0 S' -S'}_{L'_2 1 L'} I^{0 S'' -S''}_{L_3 1 L''} I^{0 S'' -S''}_{L''_2 1 L''}} \nonumber \\
&&\qquad\quad =  
\begin{cases}
8 I_{L'_3 1 L}^{0 1 -1} I_{L_2 1 L}^{0 1 -1} I_{L''_3 1 L'}^{0 1 -1} I_{L'_2 1 L'}^{0 1 -1} I_{L_3 1 L''}^{0 1 -1} I_{L''_2 1 L''}^{0 1 -1} & {\rm for \ } L_3' + L_2, L_3'' + L_2', L_3 + L_2'' = {\rm even}  \\
0 & {\rm otherwise}
\end{cases}~.
\end{eqnarray}
By the same token, the summations over $\lambda_1, \lambda_2$ and $\lambda_3$
(green part in Eq.~(\ref{eq:fullbispectrum})) are given by
\begin{eqnarray}
&&\OliveGreen{ \sum_{\lambda_1, \lambda_2, \lambda_3 = \pm 1}
I^{0 \lambda_1 -\lambda_1}_{L_1 \ell_1 2} 
I^{0 \lambda_2 -\lambda_2}_{L'_1 \ell_2 2} 
I^{0 \lambda_3 -\lambda_3}_{L''_1 \ell_3 2} } 
=
\begin{cases}
8 I_{L_1 \ell_1 2}^{0 1 -1} I_{L_1' \ell_2 2}^{0 1 -1} I_{L_1'' \ell_3 2}^{0 1 -1}  &
{\rm for \ } L_1 + \ell_1, L_1' + \ell_2, L_1'' + \ell_3 = {\rm even}  \\
0 & {\rm otherwise}
\end{cases} ~.
\end{eqnarray} 
Then, using the function ${\cal K}^N_{ll'}$ and the above equations,
the CMB bispectrum of Eq. (\ref{eq:CMB_vec_bis}) can be written as
\begin{eqnarray}
B^{(V V V)}_{I I I, \ell_1 \ell_2 \ell_3} 
&\simeq& \left( - \frac{32 (2\pi)^{1/2}}{3 \rho_{\gamma, 0}} \right)^3  
\left[ \prod\limits^3_{n=1}
4\pi
\int_0^\infty {k_n^2 dk_n \over (2\pi)^3}
\mathcal{T}^{(V)}_{I, \ell_n}(k_n)
j_{\ell_n} (k_n (\tau_0-\tau_*))
\right] \nonumber \\
&& \times
\sum_{L_1 L_1' L_1''} 
I_{L_1 \ell_1 2}^{0 1 -1} 
I_{L_1' \ell_2 2}^{0 1 -1} 
I_{L_1'' \ell_3 2}^{0 1 -1}
\sum_{L L' L''} 
\left\{
  \begin{array}{ccc}
  \ell_1 & \ell_2 & \ell_3 \\
  L' & L'' & L 
  \end{array}
 \right\} 
\nonumber \\
&& \times 
\sum_{\substack{L_2 L'_2 L''_2 \\ L'_3 L''_3 L_3}} 
A_B^3 (\tau_0-\tau_*)^6 \left( {\tau_* \over 5} \right)^3 
\mathcal{K}_{L_2 L'_3}^{-(n_B +1)}(\tau_0-\tau_*)
\mathcal{K}_{L'_2 L''_3}^{-(n_B +1)}(\tau_0-\tau_*)
\mathcal{K}_{L''_2 L_3}^{-(n_B +1)}(\tau_0-\tau_*) \nonumber \\
&& \times
(-1)^{\sum_{i=1}^3 \frac{\ell_i+ L_i + L'_i + L''_i}{2}}
I^{0~0~0}_{L_1 L_2 L_3} 
 I^{0~0~0}_{L'_1 L'_2 L'_3}  I^{0~0~0}_{L''_1 L''_2 L''_3} 
I_{L'_3 1 L}^{0 1 -1} I_{L_2 1 L}^{0 1 -1} 
I_{L''_3 1 L'}^{0 1 -1} I_{L'_2 1 L'}^{0 1 -1} 
I_{L_3 1 L''}^{0 1 -1} I_{L''_2 1 L''}^{0 1 -1}
\nonumber \\
&& \times
\left\{
  \begin{array}{ccc}
  L'' & L & \ell_1 \\
  L_3 & L_2 & L_1 \\
  1 & 1 & 2
  \end{array}
 \right\}
\left\{
  \begin{array}{ccc}
  L & L' & \ell_2 \\
  L'_3 & L'_2 & L'_1 \\
  1 & 1 & 2
  \end{array}
 \right\}
\left\{
  \begin{array}{ccc}
  L' & L'' & \ell_3 \\
  L''_3 & L''_2 & L''_1 \\
  1 & 1 & 2
  \end{array}
 \right\} 
  \label{eq:CMB_vec_bis_approx}~.
\end{eqnarray}
Here from the selection rules of the Wigner symbols
\cite{Shiraishi:2010kd}, we can further limit
the summation range of the multipoles as   
\begin{eqnarray}
\begin{split}
& |\ell_1 - \ell_2| \leq \ell_3 \leq \ell_1 + \ell_2 ~, \\
& L_1 = |\ell_1 \pm 2|,~ \ell_1 ~, \ \ L^\prime_1 = |\ell_2 \pm 2|,~ \ell_2 ~, \
 \ L^{\prime\prime}_1 = |\ell_3 \pm 2|,~ \ell_3 ~, \\
& |L - \ell_2| \leq L^\prime \leq L + \ell_2 ~ , \ \ 
{\rm Max}[|L - \ell_1|, |L^\prime - \ell_3|] \leq L^{\prime\prime} \leq {\rm Min}[ L +
 \ell_1, L' + \ell_3] ~ , \\
& (L_2, L'_3) =  (|L-1|, |L \pm 1|),~ (L,L),~ (L+1, |L \pm 1|)~, \\
& (L'_2, L''_3) =  (|L'-1|, |L' \pm 1|),~ (L',L'),~ (L'+1, |L' \pm 1|)~, \\
& (L''_2, L_3) = (|L''-1|, |L'' \pm 1|),~ (L'',L''),~ (L''+1, |L'' \pm
 1|)~, \\
& L_1 + L_2 + L_3 = {\rm even}~, \ \ 
L'_1 + L'_2 + L'_3 = {\rm even}~, \ \
L''_1 + L''_2 + L''_3 = {\rm even}~, \\
& |L_1 - L_2| \leq L_3 \leq L_1 + L_2 ~ , \ \ |L'_1 - L'_2| \leq  L'_3
 \leq L'_1 + L'_2 ~ , \ \ |L''_1 - L''_2| \leq  L''_3 \leq L''_1 + L''_2
~. \label{eq:range_L1Lk} 
\end{split}  
\end{eqnarray}
and from the above restrictions the multipoles in the bispectrum,
$\ell_1, \ell_2$ and $\ell_3$, are also limited as
\begin{eqnarray}
\begin{split}
& \ell_1 + \ell_2 + \ell_3 = {\rm even} ~ .
\end{split}
\end{eqnarray}
Therefore, these selection rules significantly reduce the number of
calculation.  In these ranges, while $L'$ and $L''$ are limited by
$L$, only $L$ has no upper bound. However, we can show that
the summation of $L$ is suppressed at $\ell_1\sim \ell_2\sim \ell_3 \ll
L$ as follows.
When the summations
with respect to $L,L'$ and $L''$ are evaluated at large $L, L'$ and
$L''$, namely $\ell_1, \ell_2, \ell_3 \ll L \sim L' \sim L'', L_2
\sim L'_3 \sim L, L'_2 \sim L''_3 \sim L'$ and $L''_2 \sim L_3 \sim
L''$, we get
\begin{eqnarray}
&&\sum_{L L' L''}
 \left\{
  \begin{array}{ccc}
  \ell_1 & \ell_2 & \ell_3 \\
  L' & L'' & L 
  \end{array}
 \right\} 
\sum_{\substack{L_2 L'_2 L''_2 \\ L'_3 L''_3 L_3}} 
\mathcal{K}_{L_2 L'_3}^{-(n_B +1)}(\tau_0-\tau_*)
\mathcal{K}_{L'_2 L''_3}^{-(n_B +1)}(\tau_0-\tau_*)
\mathcal{K}_{L''_2 L_3}^{-(n_B +1)}(\tau_0-\tau_*) \nonumber \\
&& \qquad\qquad \times (-1)^{\sum_{i=1}^3 \frac{L_i + L_i' + L''_i}{2}} 
 I^{0~0~0}_{L_1 L_2 L_3} 
 I^{0~0~0}_{L'_1 L'_2 L'_3}  I^{0~0~0}_{L''_1 L''_2 L''_3} 
I_{L'_3 1 L}^{0 1 -1} I_{L_2 1 L}^{0 1 -1} 
I_{L''_3 1 L'}^{0 1 -1} I_{L'_2 1 L'}^{0 1 -1} 
I_{L_3 1 L''}^{0 1 -1} I_{L''_2 1 L''}^{0 1 -1}
\nonumber \\
&& \qquad\qquad \times 
\left\{
  \begin{array}{ccc}
  L'' & L & \ell_1 \\
  L_3 & L_2 & L_1 \\
  1 & 1 & 2
  \end{array}
 \right\}
\left\{
  \begin{array}{ccc}
  L & L' & \ell_2 \\
  L'_3 & L'_2 & L'_1 \\
  1 & 1 & 2
  \end{array}
 \right\}
\left\{
  \begin{array}{ccc}
  L' & L'' & \ell_3 \\
  L''_3 & L''_2 & L''_1 \\
  1 & 1 & 2
  \end{array}
 \right\} \nonumber \\
&&\qquad \propto \sum_{L L' L''} (L L' L'')^{n_B + 4/3} 
 ~.
\end{eqnarray}
Therefore, we may obtain a stable result with the summations
over a limited number of $L$ when we consider the magnetic power
spectrum is as red as $n_B \sim -2.9$, because the summations of $L^\prime$ and
$L^{\prime\prime}$ are limited by $L$. Here, we use the analytic
formulas of the $I$ symbols which are given by
\begin{eqnarray} 
\left\{
  \begin{array}{ccc}
  \ell_1 & \ell_2 & \ell_3 \\
  L' & L'' & L 
  \end{array}
 \right\} \propto (L L' L'')^{-1/6}~, \ \
{\cal K}_{L_2 L'_3}^{-(n_B + 1)} \propto L^{n_B+1}~, \ \
\left\{
  \begin{array}{ccc}
  L'' & L & \ell_1 \\
  L_3 & L_2 & L_1 \\
  1 & 1 & 2
  \end{array}
 \right\} \propto (L'' L)^{-1/2}~,
\end{eqnarray}
as described in detail in Appendix \ref{appen:wigner}.

Using the approximation and the summation rules described above, we can
perform the computation of the CMB bispectrum containing full-angular
dependence in a reasonable time.

\section{Results}
\label{sec:result}

Now we show the result of the CMB temperature bispectrum induced from
the vector anisotropic stress $\Pi_{Bv}^{(\lambda)}$.  In order to
compute numerically, we insert Eq. (\ref{eq:CMB_vec_bis_approx}) into
the Boltzmann code for anisotropies in the microwave background (CAMB)
\cite{Lewis:2004ef, Lewis:1999bs}.  We use the transfer function of
magnetic-compensated modes calculated as Refs. \cite{Shaw:2009nf,
Lewis:2004sec}, which is shown in Appendix~\ref{appen:transfer}. 
In the calculation of the Wigner-$3j, 6j$ and 9j symbols, we use a
common mathematical library called SLATEC \cite{slatec} and analytical
expressions in Appendix \ref{appen:wigner}.

In Fig.~\ref{fig:vec_III_samel_difnB.eps}, we show the reduced bispectra
of the temperature fluctuation induced by the PMFs defined
as~\cite{Komatsu:2001rj}
\begin{eqnarray}
b^{(VVV)}_{III,\ell_1 \ell_2 \ell_3}
\equiv \left( I^{0~0~0}_{\ell_1 \ell_2 \ell_3} \right)^{-1} 
B^{(VVV)}_{III,\ell_1 \ell_2 \ell_3}~,
\end{eqnarray} 
for $\ell_1 = \ell_2 = \ell_3$.  Red solid and green dashed lines
correspond to the bispectrum given by Eq.~(\ref{eq:CMB_vec_bis_approx})
with the spectral index of the power spectrum of PMFs fixed as $n_B = -
2.9$ and $- 2.8$, respectively.  One can see that the peak of each
bispectrum is located at $\ell \sim 1500$ and the position is similar to
that of the angular power spectrum $C^{(V)}_{I, \ell}$ induced from the
vector mode as calculated in Appendix \ref{appen:power}.  At small
scales, the vector mode contributes to the CMB power spectrum through
the Doppler effect.  Thus we can easily find that through this Doppler
effect the vector mode can also enhance the CMB bispectrum.  In our
related paper \cite{Shiraishi:2010yk}, we have also shown that the
contribution from the vector mode dominates over that from the scalar mode
at the scales around $\ell \sim 1500$ in the CMB bispectrum induced from
the PMFs, as in the CMB power spectrum.

As for the amplitude of the CMB bispectrum of the vector mode induced
from the PMFs, one can expect $b_{III,\ell \ell \ell}^{(VVV)} \sim
C_{I,\ell}^{(V) 3/2}$ by using the amplitude of the CMB power spectrum
of the vector mode induced from the PMFs.  However, in
Fig.~\ref{fig:vec_III_samel_difnB.eps} we find that the amplitude of
$b_{III, \ell \ell \ell}^{(VVV)}$ is smaller than the above expectation.  This is
because the configuration of multipoles, corresponding to the angles of
wave number vectors, is limited to the conditions placed by the Wigner
symbols.  

We can understand this by considering the scaling relation
with respect to $\ell$.  If the magnetic power spectrum given by
Eq. (\ref{eq:def_power}) is close to the scale-invariant shape, the
configuration that satisfies $L \sim L'' \sim \ell$ and $L' \sim 1$
contributes dominantly in the summations.  Furthermore, the other
multipoles are evaluated as 
\begin{eqnarray}
L_1 \sim L'_1 \sim L''_1 \sim \ell~, \ \
L_2 \sim L''_2 \sim L_3 \sim L'_3  \sim \ell~, \ \ L'_2 \sim L''_3 \sim 1~,
\end{eqnarray}
from the triangle conditions described in Appendix
\ref{appen:wigner}. Then we can find $b^{(VVV)}_{III, \ell \ell \ell}
\propto \ell^{2 n_B + 4}$ for $\ell \lesssim 1000$, where we have also
used the following relations
\begin{eqnarray}
\begin{split}
& \int k^2 dk \mathcal{T}^{(V)}_{I, \ell_i}(k) j_{\ell_i} (k(\tau_0 - \tau_*))
 \propto \ell~, \ \ 
\left\{
  \begin{array}{ccc}
  \ell_1 & \ell_2 & \ell_3 \\
  L' & L'' & L 
  \end{array}
 \right\} 
\propto \ell^{-1}~, \ \
{\cal K}_{L_2 L'_3}^{-(n_B + 1)} \sim {\cal K}_{L''_2 L_3}^{-(n_B + 1)} 
 \propto \ell^{n_B+1}~, \\
& \left\{
  \begin{array}{ccc}
  L'' & L & \ell_1 \\
  L_3 & L_2 & L_1 \\
  1 & 1 & 2
  \end{array}
 \right\} 
\propto \ell^{-3/2}~, \ \ 
\left\{
\begin{array}{ccc}
  L & L' & \ell_1 \\
  L'_3 & L'_2 & L_1' \\
  1 & 1 & 2
  \end{array}
 \right\} \sim 
\left\{
\begin{array}{ccc}
  L' & L'' & \ell_1 \\
  L''_3 & L''_2 & L''_1 \\
  1 & 1 & 2
  \end{array}
 \right\} 
\propto \ell^{-1}~,
\end{split}
\end{eqnarray}
which, except for the first relation, are also coming from the
triangle conditions of the Wigner $3$-j symbols.  Therefore, combining
with the scaling relation of the CMB power spectrum mentioned in Appendix
\ref{appen:power}, we find that $b^{(VVV)}_{III, \ell \ell \ell}$ is
suppressed by a factor $\ell^{(n_B - 1)/2}$ from $C^{(V) 3/2}_{I,\ell}$.
This is the reason why our constraint on the PMF from the vector
bispectrum is not so much stronger than expected from the scalar
counterpart.

From this figure, we also find that the CMB bispectrum becomes steeper
if $n_B$ becomes larger, which is similar to the case of the power
spectrum.  This will lead to another constraint on the strength of the
PMFs.  In particular, as shown in
Refs.~\cite{Shiraishi:2010yk,Brown:2005kr,Seshadri:2009sy,Caprini:2009vk,Cai:2010uw,
Trivedi:2010gi}, although the CMB bispectrum induced from the PMFs is
dominated by the contribution from the scalar mode on large scales, such
contribution becomes small on small scales.  Therefore, it will be
important to consider not only the contribution from the scalar mode
induced from the PMFs on large scales but also that from the vector mode
on small scales to obtain the constraint on the amplitude and the
spectral index of the PMFs' power spectrum simultaneously.

\begin{figure}[t]
  \begin{center}
    \includegraphics{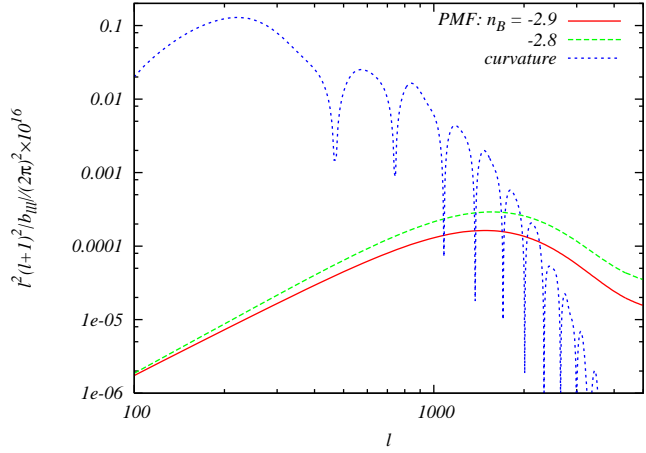}
  \end{center}
  \caption{(color online). 
Absolute values of the normalized reduced bispectra of temperature fluctuation for a configuration $\ell_1 = \ell_2 = \ell_3$. The lines correspond to the spectra generated from vector anisotropic stress for $n_B = -2.9$ (red solid line) and $-2.8$ (green dashed line), and primordial non-Gaussianity with $f^{\rm local}_{\rm NL} = 5$ (blue dotted line). The strength of PMFs is fixed to $B_{1 {\rm Mpc}} = 4.7 {\rm nG}$ and the other cosmological parameters are fixed to the mean values limited from WMAP-7yr data reported in Ref. \cite{Komatsu:2010fb}.}
  \label{fig:vec_III_samel_difnB.eps}
\end{figure}

\begin{figure}[t]
  \begin{center}
    \includegraphics{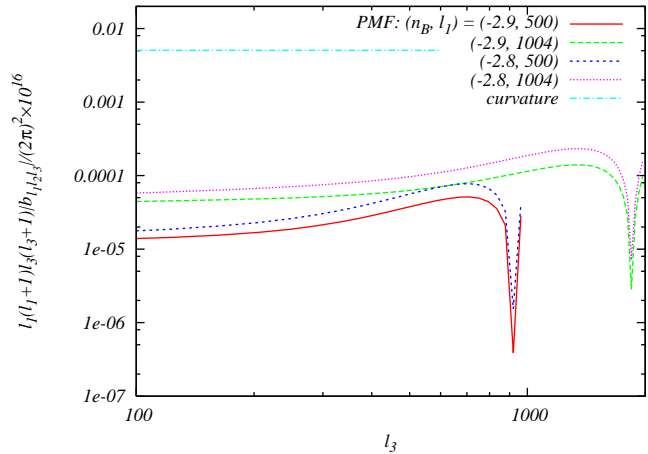}
  \end{center}
  \caption{(color online). Absolute values of the normalized reduced
 bispectra of temperature fluctuation given by Eq.~(\ref{eq:CMB_vec_bis_approx}) and generated by primordial
  non-Gaussianity given by Eq.~(\ref{eq:cmb_bis_local}) as a function of $\ell_3$ with $\ell_1$ and $\ell_2$ fixed to some value as indicated.
Each parameter is fixed to the same value defined in Fig. \ref{fig:vec_III_samel_difnB.eps}.}
  \label{fig:vec_III_difl_difnB.eps}
\end{figure}

In Fig.~\ref{fig:vec_III_difl_difnB.eps}, we show the reduced bispectrum
$b_{III, \ell_1 \ell_2 \ell_3}$ with respect to $\ell_3$ with setting
$\ell_1 = \ell_2$.  From this figure we can see that the normalized
reduced bispectrum of the vector mode induced from the PMFs for $\ell_1,
\ell_2, \ell_3 \gtrsim 100$ is nearly flat and given as
\begin{eqnarray}
\ell_1 (\ell_1+1) \ell_3 (\ell_3+1) |b^{(VVV)}_{III, \ell_1 \ell_2 \ell_3}| 
\sim 2 \times 10^{-19} 
\left(\frac{B_{1 \rm Mpc}}{4.7 \rm nG} \right)^6 . \nonumber \\ 
\label{eq:CMB_vec_bis_rough}
\end{eqnarray}
It is seen that $b^{(VVV)}_{III, \ell_1 \ell_2 \ell_3}$ for $n_B\simeq
-3$ dominates in $\ell_1 = \ell_2 \gg \ell_3$. This means that the shape
of the CMB bispectrum generated from the vector anisotropic stress of
the PMF is close to the so-called local-type configuration if the power
spectrum of the PMF is nearly scale invariant. We can understand this by
the analytical evaluation as follows. As mentioned above, in the
summations of Eq. (\ref{eq:CMB_vec_bis_approx}), the configuration that
$L \sim \ell_1, L' \sim 1$ and $L'' \sim \ell_3$ contributes dominantly.
By using this and the approximations that 
\begin{eqnarray}
L_1 \sim \ell_1~, \ \ L'_1 \sim \ell_2~, \ \ L''_1 \sim \ell_3~, \ \
L_2 \sim L'_3 \sim L~, \ \ L'_2 \sim L''_3 \sim L'~, \ \ L''_2 \sim L_3 \sim L''~,
\end{eqnarray}
which again come from the triangle conditions from the Wigner symbols,
the scaling relation of $\ell_3$ at large scale is evaluated as
$b^{(VVV)}_{III,\ell_1 \ell_2 \ell_3} \propto \ell_3^{n_B + 1}$.  From
this estimation we can find that $\ell_1(\ell_1+1)\ell_3(\ell_3 +
1)b^{(VVV)}_{III, \ell_1 \ell_2 \ell_3} \propto \ell_3^{0.1}$, for $n_B
= -2.9$, and $\ell_3^{0.2}$ for $n_B = -2.8$, respectively, which match
the behaviors of the bispectra in
Fig. ~\ref{fig:vec_III_difl_difnB.eps}.

In order to obtain a valid constraint on the magnitude of the PMF, we
compare the bispectrum induced from the PMF with that from the
local-type primordial non-Gaussianity in the curvature perturbations,
which is typically estimated as \cite{Riot:2008ng}
\begin{eqnarray}
\ell_1 (\ell_1+1) \ell_3
(\ell_3+1)b_{\ell_1 \ell_2 \ell_3} \sim 4 \times 10^{-18} f^{\rm
local}_{\rm NL}~. \label{eq:cmb_bis_local}
\end{eqnarray}
By comparing this with
Eq. (\ref{eq:CMB_vec_bis_rough}), the relation between the magnitudes of
the PMF with the nearly scale-invariant power spectrum and $f^{\rm
local}_{\rm NL}$ is derived as
\begin{eqnarray}
\left( \frac{B_{1\rm Mpc}}{1\rm nG} \right) \sim 7.74~ 
|f^{{\rm local}}_{\rm NL}|^{1/6} \ \ ({\rm for ~ n_B \sim -3})~.
\end{eqnarray} 

By making use of the above relation, we can place the upper bound of
strength of the PMF.  If we assume $|f^{\rm local}_{\rm NL}| < 100$ as
considered in Ref. \cite{Seshadri:2009sy}, we can translate this to the
constraint on the PMF amplitude as $B_{1\rm Mpc} <
17{\rm nG}$, which is stronger by a factor of 2 than estimated in
Ref. \cite{Seshadri:2009sy}.
On the other hand, from the current observational lower bound from WMAP 7-yr data mentioned in Sec. \ref{sec:intro},
namely $f_{\rm NL}^{\rm local} > -10$, we derive $B_{1\rm Mpc} < 11{\rm nG}$.
If we use $|f_{\rm NL}^{\rm local}| < 5$ which is expected
from Planck experiment \cite{:2006uk}, we will
meet a tight constraint as $B_{1\rm Mpc} < 10 {\rm nG}$.

\section{Summary and Discussion}
\label{sec:sum}

In this paper, we present a calculation method of the bispectrum of CMB
temperature fluctuation induced from the vector mode of the PMFs as
described in Ref. \cite{Shiraishi:2010sm}, by taking into account the
full angular dependence of the bispectrum of magnetic fields. We expand
all the angular dependence with the spin spherical harmonics and convert
them to the summations of the Wigner symbols.  In the radial integrals
and timelike integrals, we use only one approximation that the interval
of the timelike integrals is confined to the moment of the
recombination, which corresponds to neglecting the vector mode ISW effect.
This approximation is valid because the radiation transfer function of the vector magnetic mode has a sharp peak around $k \sim \ell/(\tau_0
- \tau_*)$ which comes from the Doppler effect of the baryon vorticity
induced from the magnetic field.  We checked that the errors by the
approximation are less than $10 \%$.

As the results, it is found that the CMB bispectrum from the magnetic
vector mode dominates at small scales compared to that from the magnetic
scalar mode which has been calculated in the literature. It is also
found that the bispectrum has significant signals on the squeezed limit,
namely the local-type configuration, if the magnetic field power
spectrum is nearly scale invariant. This is understood by considering
the asymptotic scaling relation of the CMB bispectrum.  We
also investigate the dependence of the spectral index of the power
spectrum of the PMFs on the CMB bispectrum and we find that the CMB
bispectrum of the vector mode induced from the PMFs is more sensitive to
the spectral index of the PMFs' power spectrum than that of the scalar
mode.  Hence, we conclude that it is important to consider not only
the contribution from the scalar mode of the PMFs on large
scales, but also that from the vector mode on small scales to obtain
the constraint on the amplitude and the spectral index of the PMFs'
power spectrum simultaneously.

By translating the current bound on the local-type non-Gaussianity from
the CMB bispectrum into the bound on the amplitude of the magnetic
fields, we obtain a new limit: $B_{1\rm Mpc} < 11 {\rm nG}$. This is a
rough estimate and a tighter constraint is expected if one considers the
full $\ell$ contribution by using an appropriate estimator of the CMB
bispectrum induced from the primordial magnetic fields.

Because of the complicated discussions and mathematical manipulations, here
we restrict our attention to the temperature bispectrum from the vector
mode of the PMFs. However, one will be able to apply this methodology to
the bispectra of CMB temperature and polarization from the scalar,
vector and tensor modes.

\begin{acknowledgments}
We would like to thank Dai G. Yamazaki for useful discussion and Tina Kahniashvili for private communication. 
This work is supported by Grant-in-Aid for JSPS Research under Grant
 No. 22-7477 (M. S.), JSPS Grant-in-Aid for Scientific
Research under Grant Nos. 22340056 (S. Y.), 21740177, 22012004 (K. I.),
 and 21840028 (K. T.).
This work is supported in part by the Grant-in-Aid for Scientific
Research on Priority Areas No. 467 "Probing the Dark Energy through an
 Extremely Wide and Deep Survey with Subaru Telescope" and by the
 Grant-in-Aid for Nagoya University Global COE Program, "Quest for
 Fundamental Principles in the Universe: from Particles to the Solar
 System and the Cosmos," from the Ministry of Education, Culture,
 Sports, Science and Technology of Japan. 
\end{acknowledgments}

\appendix


\section{CMB temperature fluctuations induced from vector anisotropic stresses}\label{appen:transfer}

In Refs.~\cite{Mack:2001gc, Kahniashvili:2010us}, it is discussed that the temperature fluctuations are generated via Doppler and integrated Sachs-Wolfe effects on the CMB vector modes. Based on them, we derive the transfer function of the vector magnetic mode as follows.

When we decompose the metric perturbations into vector components as
\begin{eqnarray}
\delta g_{0c} &=& \delta g_{c0} = a^2 A_c ~, \\
\delta g_{cd} &=& a^2 (\partial_c h^{(V)}_{d} + \partial_d h^{(V)}_{c})~,
\end{eqnarray}
we can construct two gauge-invariant variables, namely a vector perturbation of the extrinsic curvature and a vorticity, as
\begin{eqnarray}
{\bf V} &\equiv& {\bf A} - {\bf h}' ~, \\
{\bf \Omega} &\equiv& {\bf v} - {\bf A} ~,
\end{eqnarray}
where ${\bf v}$ is the spatial part of the four-velocity perturbation of
a stationary fluid element and a dash denotes a partial derivative of the conformal time $\tau$.
Here, choosing a gauge as ${\bf h}' = 0$, we can express the Einstein equation
\begin{eqnarray}
{\bf V}' + 2 \frac{a'}{a} {\bf V} = - \frac{16 \pi G \rho_{\gamma,0} ({\bf \Pi_\gamma^{(V)}} + {\bf \Pi_\nu^{(V)}} + {\bf \Pi_B^{(V)}})}{a^2 k} ~, \label{eq:einstein}
\end{eqnarray}
and the Euler equations for photons and baryons
\begin{eqnarray}
{\bf \Omega_\gamma}' + \tau_c'({\bf v_\gamma} - {\bf v_b}) 
&=& 0 ~, \label{eq:euler_g}\\
{\bf \Omega_b}' + \frac{a'}{a} {\bf \Omega_b} - \frac{\tau_c'}{R}({\bf v_\gamma} - {\bf v_b}) 
&=& \frac{k \rho_{\gamma,0}{\bf \Pi_B^{(V)}}}{a^4 (\rho_b + p_b)}~. \label{eq:euler_b}
\end{eqnarray}
Here $\Pi^{(V)}_a = -i \hat{k}_b P_{ac} \Pi_{bc}$, $p$ is the isotropic
pressure, the indices $\gamma, \nu$ and $b$ denote the photon, neutrino and baryon, $\tau_c$ is the optical depth, and $R \equiv (\rho_b + p_b)/(\rho_\gamma + p_\gamma)$.
In the tight-coupling limit as ${\bf v_\gamma} \simeq {\bf v_b}$, the photon vorticity is comparable to the baryon one: ${\bf \Omega_\gamma} \simeq {\bf \Omega_b} \equiv {\bf \Omega}$. Then, the Euler equations (\ref{eq:euler_g}) and (\ref{eq:euler_b}) are combined into
\begin{eqnarray}
(1+R){\bf \Omega}' + R \frac{a'}{a} {\bf \Omega} 
= \frac{k \rho_{\gamma,0}{\bf \Pi_B^{(V)}}}{a^4 (\rho_\gamma + p_\gamma)}~,  
\end{eqnarray}
and this solution is given by 
\begin{eqnarray}
{\bf \Omega}({\bf k},\tau) &\simeq& \beta(k,\tau) {\bf \Pi_B^{(V)}(k)} ~, \\ 
\beta(k,\tau) &=& 
\begin{cases}
\frac{k \tau \rho_{\gamma,0}}{(1+R) (\rho_{\gamma,0} + p_{\gamma,0})} & {\rm for} \ k<k_S \\
\frac{5 \tau_c' \rho_{\gamma,0}}{k(\rho_{\gamma,0} + p_{\gamma,0})} & {\rm for} \ k>k_S
\end{cases}~,
\end{eqnarray}
where $k_S$ means the Silk damping scale.

As mentioned above, the CMB temperature anisotropies of vector modes are produced through Doppler and integrated Sachs-Wolfe effect as 
\begin{eqnarray}
\frac{\Delta T (\hat{\bf n})}{T}
&=& -{\bf v_\gamma} \cdot \hat{\bf n} |^{\tau_0}_{\tau_*} + \int_{\tau_*}^{\tau_0} d \tau {\bf V}' \cdot \hat{\bf n} ~, 
\end{eqnarray}
where $\tau_0$ is today and $\tau_*$ is the recombination epoch in
conformal time, $\mu_{k,n} \equiv \hat{\bf k} \cdot \hat{\bf n}$, $x
\equiv k(\tau_0 - \tau)$, and $\hat{\bf n}$ is an unit vector along the
line-of-sight direction. Because of compensation of the anisotropic
stresses, a solution of the Einstein equation (\ref{eq:einstein})
expresses the decaying signature as ${\bf V} \propto a^{-2}$ after
neutrino decoupling. Therefore, in an integrated Sachs-Wolfe effect
term, the contribution around the recombination epoch is dominant. Furthermore, neglecting dipole contribution due to ${\bf v}$ today, we can form the coefficient of anisotropies as 
\begin{eqnarray}
a_{\ell m} &\equiv&  \int d^2 \hat{\bf n}  \frac{\Delta T(\hat{\bf n})}{T} Y^*_{\ell m} (\hat{\bf n}) \nonumber \\
&\simeq& \int \frac{d^3 {\bf k}}{(2 \pi)^3}
\int d^2 \hat{\bf n}
[{\bf \Pi_B^{(V)}}({\bf k}) \cdot \hat{\bf n}]  Y^*_{\ell m} (\hat{\bf n}) \nonumber \beta(k,\tau_*) e^{- i \mu_{k,n} x_*} ~. 
\end{eqnarray}
In the transformation $\hat{\bf n} \rightarrow (\mu_{k,n}, \phi_{k,n})$,
the functions are rewritten as  
\begin{eqnarray}
{\bf \Pi_B^{(V)} (\bf k)} \cdot \hat{\bf n}
&\rightarrow& -i \sqrt{\frac{1- \mu_{k,n}^2}{2}} 
\sum_{\lambda = \pm 1} \Pi^{(\lambda)}_{Bv}({\bf k}) e^{i \lambda \phi_{k,n}} 
~, \\
Y^*_{\ell m}(\hat{\bf n}) &\rightarrow& \sum_{m'} D^{(\ell)}_{m m'} 
\left( S(\hat{\bf k}) \right) Y^*_{\ell m'}(\Omega_{k,n}) ~, \\
d^2 \hat{\bf n} &\rightarrow& d\Omega_{k,n}~,
\end{eqnarray}
where we use the relation: 
$\displaystyle \Pi^{(V)}_a = \sum_{\lambda = \pm 1} - i \Pi_{Bv}^{(\lambda)} \epsilon_a^{(\lambda)}$ and the Wigner $D$ matrix under the rotational transformation of an unit vector parallel to $z$ axis into $\hat{\bf k}$ corresponding to Eq.~(A7) of Ref.~\cite{Shiraishi:2010sm}. 
Therefore, performing the integration over $\Omega_{k,n}$ in the same manner as Ref.~\cite{Shiraishi:2010sm}
\footnote{In Ref.~\cite{Shiraishi:2010sm}, there are three typos: right-hand sides of Eqs.~(B21), (B22) and (B23) must be multiplied by a factor $-1$, respectively.}, 
we can obtain the explicit form of $a_{\ell m}$ and express the radiation transfer function introduced in Eq.~(\ref{eq:alm_general}) as
\begin{eqnarray}
{\cal T}^{(V)}_{I,\ell}(k) 
\simeq \left[ \frac{(\ell+1)!}{(\ell-1)!} \right]^{1/2} 
\frac{\beta(k, \tau_*)}{\sqrt{2}} \frac{j_\ell(x_*)}{x_*}~.
\end{eqnarray}
This is consistent with the results presented in Refs.~\cite{Lewis:2004ef, Lewis:2004sec}.

\section{Analytic expressions of the Wigner symbols}\label{appen:wigner}

The Wigner-$3j, 6j$ and $9j$ symbols express Clebsch-Gordan coefficients
between two other eigenstates coupled to two three, and four individual
momenta \cite{Hu:2001fa, Gurau:2008, Jahn/Hope:1954}. Their selection
rules and several properties are reviewed in Ref.~\cite{mathematica,
Shiraishi:2010kd}. Here, using their knowledge, we show the
analytical formulas of the Wigner symbols which appear in the CMB
bispectrum of Eq.~(\ref{eq:CMB_vec_bis_approx}).

The $I$ symbols, which are defined as 
$I^{s_1 s_2 s_3}_{l_1 l_2 l_3}
\equiv \sqrt{\frac{(2 l_1 + 1)(2 l_2 + 1)(2 l_3 + 1)}{4 \pi}}
\left(
  \begin{array}{ccc}
  l_1 & l_2 & l_3 \\
  s_1 & s_2 & s_3
  \end{array}
 \right)$, are expressed as 
\begin{eqnarray}
I_{l_1 l_2 l_3}^{0~0~0} 
&=& 
\sqrt{\frac{\prod_{i=1}^3 (2 l_i+1)}{4 \pi}} 
(-1)^{\sum_{i=1}^3 \frac{- l_i}{2}} \nonumber \\
&&\times 
\frac{ \left(\sum_{i=1}^3 \frac{l_i}{2} \right)! \sqrt{(-l_1 + l_2 + l_3)!}
\sqrt{(l_1 - l_2 + l_3)!} \sqrt{(l_1 + l_2 - l_3)!} }
{ \left(\frac{- l_1 + l_2 + l_3}{2}\right)! 
\left(\frac{l_1 - l_2 + l_3}{2}\right)! 
\left(\frac{l_1 + l_2 - l_3}{2}\right)! 
\sqrt{ (\sum_{i=1}^3 l_i + 1)!} } 
\ \   ({\rm for \ } l_1 + l_2 + l_3 = {\rm even} ) \\
&=& 0 \ \ ({\rm for \ } l_1 + l_2 + l_3 = {\rm odd} )
~, \\
I_{l_1 ~ l_2 ~ l_3}^{0~ 1~ -1} 
&=& \sqrt{\frac{5}{8\pi}} (-1)^{l_2+1}
 \sqrt{\frac{(l_2-1)(l_2+1)}{l_2 - 1/2}} \ \
({\rm for \ } l_1 = l_2 - 2, l_3 = 2) \\
&=& \sqrt{\frac{15}{16\pi}} (-1)^{l_2}
\sqrt{\frac{l_2 + 1/2}{(l_2 - 1/2)(l_2 + 3/2)}} \ \ 
({\rm for \ } l_1 = l_2, l_3 = 2) \\
&=& \sqrt{\frac{5}{8\pi}} (-1)^{l_2} \sqrt{\frac{l_2(l_2+2)}{l_2 + 3/2}}
 \ \
({\rm for \ } l_1 = l_2 + 2, l_3 = 2) \\
&=& \sqrt{\frac{3}{8\pi}} (-1)^{l_3+1}
\sqrt{l_3 + 1} \ \
({\rm for \ } l_1 = l_3 - 1, l_2 = 1) \\
&=& \sqrt{\frac{3}{4\pi}} (-1)^{l_3+1} \sqrt{l_3 + 1/2} \ \
({\rm for \ } l_1 = l_3, l_2 = 1) \\
&=& \sqrt{\frac{3}{8\pi}} (-1)^{l_3+1}
\sqrt{l_3} \ \
({\rm for \ } l_1 = l_3 + 1, l_2 = 1)~.
\end{eqnarray}
Three Wigner-$9j$ symbols in Eq.~(\ref{eq:CMB_vec_bis_approx}) are    
calculated as 
\begin{eqnarray}
\left\{
  \begin{array}{ccc}
  l_1 & l_2 & l_3 \\
  l_4 & l_5 & l_6 \\
  1 & 1 & 2 
  \end{array}
 \right\}
&=& 
\sqrt{\frac{2(l_3 \pm 1) + 1}{5}}
\left\{
  \begin{array}{ccc}
  l_1 & l_4 & 1 \\
  l_3 \pm 2 & l_3 \pm 1 & l_5 
  \end{array}
 \right\}
\left\{
  \begin{array}{ccc}
  l_2 & l_5 & 1 \\
  l_3 \pm 1 & l_3 & l_1 
  \end{array}
 \right\} \ \
({\rm for \ } l_6 = l_3 \pm 2 ) \\
&=&
\sqrt{\frac{(2 l_3 - 1)(2 l_3 + 2)(2 l_3 + 3)}{30 (2 l_3)(2 l_3 + 1)}}
\left\{
  \begin{array}{ccc}
  l_1 & l_4 & 1 \\
  l_3 & l_3 - 1 & l_5 
  \end{array}
 \right\}
\left\{
  \begin{array}{ccc}
  l_2 & l_5 & 1 \\
  l_3-1 & l_3 & l_1 
  \end{array}
 \right\} \nonumber \\
&& + 
\sqrt{\frac{2 (2 l_3 - 1)(2 l_3 + 1)(2 l_3 + 3)}{15 (2 l_3)(2 l_3 + 2)}}
\left\{
  \begin{array}{ccc}
  l_1 & l_4 & 1 \\
  l_3 & l_3 & l_5 
  \end{array}
 \right\}
\left\{
  \begin{array}{ccc}
  l_2 & l_5 & 1 \\
  l_3 & l_3 & l_1 
  \end{array}
 \right\} \nonumber \\
&& + 
\sqrt{\frac{(2 l_3 - 1)(2 l_3)(2 l_3 + 3)}{30 (2 l_3 + 1)(2 l_3 + 2)}}
\left\{
  \begin{array}{ccc}
  l_1 & l_4 & 1 \\
  l_3 & l_3 + 1 & l_5 
  \end{array}
 \right\}
\left\{
  \begin{array}{ccc}
  l_2 & l_5 & 1 \\
  l_3 + 1 & l_3 & l_1 
  \end{array}
 \right\}
\ \ ({\rm for \ } l_6 = l_3 ) ~,
\end{eqnarray} 
where these Wigner-$6j$ symbols are analytically given by
\begin{eqnarray}
\left\{
  \begin{array}{ccc}
  l_1 & l_2 & 1 \\
  l_4 & l_5 & l_6 
  \end{array}
 \right\}
 &=& (-1)^{l_1+l_4+l_6+1} 
\sqrt{ \frac{ {}_{l_1+l_4+l_6+2}P_{2} ~ {}_{l_1+l_4-l_6+1}P_2 }
{ {}_{2 l_4 +3}P_3 ~ {}_{2 l_1 + 1}P_3 } } \ \ ({\rm for \ } l_2 = l_1 - 1, l_5 = l_4 + 1 ) \\
 &=& (-1)^{l_1+l_4+l_6+1} 
\sqrt{ \frac{2 (l_1 + l_4 + l_6 +2)(l_1 + l_4 - l_6 + 1) (-l_1 + l_4 + l_6 + 1) (l_1 - l_4 + l_6)}
{ {}_{2 l_4 +3}P_3 ~ {}_{2 l_1 + 2}P_3 } } \nonumber \\
&&\qquad\qquad\qquad\qquad\qquad\qquad\qquad\qquad\qquad\qquad\qquad
 ({\rm for \ } l_2 = l_1, l_5 = l_4 + 1 ) \\
&=& (-1)^{l_1+l_4+l_6+1} 
\sqrt{ \frac{ {}_{-l_1+l_4+l_6+1}P_{2} ~ {}_{l_1-l_4+l_6+1}P_2 }
{ {}_{2 l_4 +3}P_3 ~ {}_{2 l_1 + 3}P_3 } } \ \ ({\rm for \ } l_2 = l_1 +
1, l_5 = l_4 + 1 ) \\
&=& (-1)^{l_1+l_4+l_6+1} 
\left[ l_4(l_4+1) + l_1(l_1-1)(l_4+1) - l_6(l_6+1) - l_1(l_1+1)l_4
\right] \nonumber \\
&&\times \sqrt{\frac{2 (l_1 + l_4 + l_6 + 1)(l_1 + l_4 - l_6)}
{ (-l_1+l_4+l_6+1) (l_1 - l_4 +
l_6) {}_{2 l_4 + 2}P_3 ~ {}_{2 l_1 + 1}P_3 } } \ \ ({\rm for \ } l_2 = l_1 -
1, l_5 = l_4 ) \\
&=& 2 (-1)^{l_1+l_4+l_6+1} 
\frac{ l_4(l_4+1) + l_1(l_1+1)(l_4+1) - l_6(l_6+1) - l_1(l_1+1)l_4 }
{ \sqrt{{}_{2 l_4 + 2}P_3~{}_{2 l_1 + 2}P_3 } } \nonumber \\
&&\qquad\qquad\qquad\qquad\qquad\qquad\qquad\qquad\qquad\qquad\qquad
 ({\rm for \ } l_2 = l_1, l_5 = l_4 ) \\
&=& (-1)^{l_1+l_4+l_6+1} 
\left[ l_4(l_4+1) + (l_1+1)(l_1+2)(l_4+1) - l_6(l_6+1) - l_1(l_1+1)l_4
\right] \nonumber \\
&&\times \sqrt{\frac{2 (-l_1 + l_4 + l_6)(l_1 - l_4 + l_6 + 1)}
{(l_1+l_4+l_6+2) (l_1 + l_4 - l_6 +1) {}_{2 l_4 + 2}P_3 ~ {}_{2 l_1 +
3}P_3 } } \ \ ({\rm for \ } l_2 = l_1 + 1, l_5 = l_4 )~.
\end{eqnarray}
Using these analytical formulas, one can reduce the time
cost involved with calculating the bispectrum of Eq.~(\ref{eq:CMB_vec_bis_approx}).

\section{CMB all-sky power spectrum of vector modes generated from PMFs}\label{appen:power}

In this section, we derive CMB power spectrum of vector modes sourced from PMFs in the same manner as presented previously and check the validity of our original approach.

From Eq. (\ref{eq:alm_general}), the CMB power spectrum of the intensity
mode induced from $\Pi_{Bv, \ell m}^{(\pm 1)}$ is formulated as  
\begin{eqnarray}
\Braket{a^{(V)}_{I, \ell_1 m_1} a^{(V) *}_{I, \ell_2 m_2} }
&=& \Blue{ \left[ \prod_{n=1}^2 4\pi \int_0^\infty {k_n^2 dk_n \over
       (2\pi)^3} \mathcal{T}^{(V)}_{I, \ell_n}(k_n) \right] }
(-i)^{\ell_1} i^{\ell_2}
\OliveGreen{ \sum\limits_{\lambda_1, \lambda_2 = \pm 1}
\lambda_1 \lambda_2 }
\Braket{\Pi_{Bv,\ell_1 m_1}^{(\lambda_1)}(k_1) \Pi_{Bv,\ell_2
m_2}^{(\lambda_2) *}(k_2)} \nonumber \\
&\equiv& C^{(V)}_{I, \ell_1} \delta_{\ell_1, \ell_2} \delta_{m_1, m_2} ~.
\end{eqnarray}
Therefore, we should simplify the initial power spectrum of $\Pi_{Bv,
\ell m}^{(\pm 1)}$ as 
\begin{eqnarray}
\Braket{\Pi^{(\lambda_1)}_{Bv, \ell_1 m_1} (k_1) 
\Pi^{(\lambda_2) *}_{Bv, \ell_2 m_2} (k_2)}
&=& (- 4 \pi \rho_{\gamma,0})^{-2}
\Purple{\int d^2 {\hat{\bf k_1}} \int d^2 \hat{\bf k_2} {}_{- \lambda_1} 
Y^*_{\ell_1 m_1} (\hat{\bf k_1}) {}_{- \lambda_2} Y_{\ell_2 m_2}(\hat{\bf
k_2}) } \nonumber \\
&& \times 
\int_0^{k_D} k_1'^2 d k' {P}_B(k_1')  \int_0^{k_D} k_2'^2 d k_2' 
{P}_B(k_2') 
\Green{ \int d^2 \hat{\bf k_1'} \int d^2 \hat{\bf k_2'} \delta({\bf k_1} -
{\bf k_1'} - {\bf k_2'}) \delta({\bf k_2} -{\bf k_2'} - {\bf k_1'}) } \nonumber \\
&& \times  
\frac{1}{4} \hat{k_1}_a \epsilon_b^{(- \lambda_1)}(\hat{\bf k_1})
\hat{k_2}_c \epsilon_d^{(\lambda_2)}(\hat{\bf k_2})
\left[ P_{ad}(\hat{\bf k_1'}) P_{bc}({\hat{\bf k_2'}}) 
+ P_{ac}({\hat{\bf k_1'}}) P_{bd}({\hat{\bf k_2'}}) \right]~. \label{eq:def_pilm_power}
\end{eqnarray}

For the first part in two permutations, we calculate $\delta$-functions and the summations with respect to $a,b,c, d$:
\begin{eqnarray}
\begin{split}
\delta({\bf k_1} - {\bf k_1'} - {\bf k_2'}) 
&= 8 \int_0^\infty A^2 dA  
\sum _{\substack{L_1 L_2 L_3 \\ \Magenta{M_1 M_2 M_3}}} 
(-1)^{\frac{L_1 + 3 L_2 + 3 L_3}{2}} 
I_{L_1 L_2 L_3}^{0~0~0} 
j_{L_1} (k_1 A) j_{L_2} (k_1' A) j_{L_3} (k_2'A) \\
& \qquad \times 
\Purple{ Y_{L_1 M_1}^*(\hat{\bf k_1}) } 
\Green{ Y_{L_2 M_2}(\hat{\bf k_1'}) Y_{L_3 -M_3}^*(\hat{\bf k_2'}) } 
\Magenta{ (-1)^{M_2} \left(
  \begin{array}{ccc}
  L_1 &  L_2 & L_3 \\
   M_1 & -M_2 & -M_3
  \end{array}
 \right) }~,  \\
\delta({\bf k_2} - {\bf k_2'} - {\bf k_1'}) 
&= 8 \int_0^\infty B^2 dB 
\sum _{\substack{L_1' L_2' L_3' \\ \Orange{M_1' M_2' M_3'} }}
(-1)^{\frac{L'_1 + 3 L'_2 + 3 L'_3}{2}}
I_{L'_1 L'_2 L'_3}^{0~0~0} 
j_{L'_1} (k_2 B) j_{L'_2} (k_2'B) j_{L'_3} (k_1'B) \\
& \qquad \times 
\Purple{ Y_{L'_1 M'_1}^*(\hat{\bf k_2}) }
\Green{ Y_{L'_2 M'_2}(\hat{\bf k_2'}) Y_{L'_3 - M'_3}^*(\hat{\bf k_1'}) }
\Orange{ (-1)^{M'_2} \left(
  \begin{array}{ccc}
  L'_1 &  L'_2 & L'_3 \\
   M'_1 & -M'_2 & -M'_3
  \end{array}
 \right) }~, \\
\hat{k_1}_a \epsilon_d^{(\lambda_2)}(\hat{\bf k_2})
P_{ad}(\hat{\bf k_1'}) 
&= \sum_{\sigma = \pm1} \sum_{\Magenta{ m_a }, \Orange{ m_d } = \pm 1, 0} 
\left(\frac{4 \pi}{3}\right)^2 \lambda_2 
\Purple{ Y_{1 m_a}(\hat{\bf k_1})
{}_{\lambda_2} Y_{1 m_d}(\hat{\bf k_2}) }
\Green{ {}_{- \sigma} Y^*_{1 m_a}(\hat{\bf
 k_1'})
{}_{\sigma} Y^*_{1 m_d}(\hat{\bf k_1'}) } ~, \\
\hat{k_2}_c \epsilon_b^{(- \lambda_1)}(\hat{\bf k_1})
  P_{bc}({\hat{\bf k_2'}}) 
&=  \sum_{\sigma' = \pm 1} \sum_{\Orange{m_c}, \Magenta{m_b} = \pm 1, 0} 
\left(\frac{4 \pi}{3}\right)^2 (- \lambda_1) 
\Purple{ Y_{1 m_c}(\hat{\bf k_2}) {}_{-\lambda_1} Y_{1 m_b}(\hat{\bf k_1}) }
\Green{ {}_{-\sigma'} Y^*_{1 m_c}(\hat{\bf k_2'})
{}_{\sigma'} Y^*_{1 m_b}(\hat{\bf k_2'}) } ~,
\end{split}
\end{eqnarray}
perform the angular integrals of the spin spherical harmonics:
 \begin{eqnarray}
\begin{split}
\Green{ \int d^2 \hat{\bf k_1'} {}_{- \sigma} Y^*_{1 m_a} 
Y_{L_2 M_2} {}_\sigma Y^*_{1 m_d} Y^*_{L'_3 -M'_3} }
& = \sum_{L \Cyan{M} S} (-1)^{\sigma + \Magenta{m_a}} I_{L_3' 1 L}^{0 -\sigma -S} I_{L_2 1
L}^{0 -\sigma -S} 
\Orange{ \left(
  \begin{array}{ccc}
  L'_3 &  1 & L \\
  - M'_3 & m_d & M 
  \end{array}
 \right) }
\Magenta{ \left(
  \begin{array}{ccc}
  L_2 &  1 & L \\
   M_2 & -m_a & M 
  \end{array}
 \right) }~, \\
\Green{ \int d^2 \hat{\bf k_2'} 
{}_{-\sigma'} Y^*_{1 m_c} Y_{L'_2 M'_2} 
{}_{\sigma'} Y^*_{1 m_b} Y^*_{L_3 -M_3} }
&= \sum_{L' \Cyan{M'} S'} (-1)^{\sigma' + \Orange{m_c}} I_{L_3 1 L'}^{0 -\sigma' -S'}
I_{L'_2 1 L'}^{0 -\sigma' -S'} 
\Magenta{ \left(
  \begin{array}{ccc}
  L_3 &  1 & L' \\
   -M_3 & m_b & M' 
  \end{array}
 \right) } 
\Orange{ \left(
  \begin{array}{ccc}
  L'_2 &  1 & L' \\
   M'_2 & -m_c & M' 
  \end{array}
 \right) } ~, \\
\Purple{ \int d^2 \hat{\bf k_1} 
{}_{-\lambda_1} Y_{1 m_b} Y_{1 m_a}
{}_{-\lambda_1} Y^*_{\ell_1 m_1} Y^*_{L_1 M_1} }
&= \sum_{L_k \Magenta{M_k} S_k} I_{L_1 \ell_1 L_k}^{0 \lambda_1 -S_k} I_{1 1 L_k}^{0 \lambda_1 -S_k}
\Magenta{ \left(
  \begin{array}{ccc}
  L_1 & \ell_1 & L_k \\
  M_1 & m_1 & M_k 
  \end{array}
 \right)
\left(
  \begin{array}{ccc}
  1 &  1 & L_k \\
  m_a & m_b & M_k 
  \end{array}
 \right) } ~, \\
\Purple{ \int d^2 \hat{\bf k_2} 
{}_{\lambda_2} Y_{1 m_d} Y_{1 m_c} 
{}_{-\lambda_2} Y_{\ell_2 m_2} Y^*_{L'_1 M'_1} }
&= \sum_{L_p \Orange{M_p} S_p} (-1)^{m_2 + \lambda_2 } I_{L'_1 \ell_2 L_p}^{0
\lambda_2 -S_p} I_{1 1 L_p}^{0 \lambda_2 -S_p} 
\Orange{ \left(
  \begin{array}{ccc}
  L'_1 & \ell_2 & L_p \\
  - M'_1 & m_2 & M_p 
  \end{array}
 \right)  
\left(
  \begin{array}{ccc}
  1 &  1 & L_p \\
  -m_c & -m_d & M_p 
  \end{array}
 \right) } ~, 
\end{split}
\end{eqnarray}
sum up the Wigner-$3j$ symbols over the azimuthal quantum numbers:
\begin{eqnarray}
\begin{split}
& \Magenta{ \sum_{\substack{M_1 M_2 M_3 \\ M_k m_a m_b}} 
(-1)^{M_2+m_a} 
\left(
  \begin{array}{ccc}
  1 & 1 & L_k \\
  m_a & m_b & M_k 
  \end{array}
 \right) 
\left(
  \begin{array}{ccc}
  L_1 & L_2 & L_3 \\
  M_1 & - M_2 & - M_3 
  \end{array}
 \right)
\left(
  \begin{array}{ccc}
  L_3 & 1 & L' \\
  - M_3 & m_b & M' 
  \end{array}
 \right) 
 \left(
  \begin{array}{ccc}
  L_2 &  1 & L \\
  M_2 & -m_a & M 
  \end{array}
 \right)
 \left(
  \begin{array}{ccc}
  L_1 & \ell_1 & L_k \\
  M_1 & m_1 & M_k 
  \end{array}
 \right) } \\
& \qquad\qquad = (-1)^{\Cyan{M} + \ell_1 + L_3 + L + 1} 
\Cyan{ \left(
  \begin{array}{ccc}
  L' & L & \ell_1 \\
  M' & -M & m_1 
  \end{array}
 \right) }
\left\{
  \begin{array}{ccc}
  L' & L & \ell_1 \\
  L_3 & L_2 & L_1 \\
  1 & 1 & L_k 
  \end{array}
 \right\}  ~, \\
& \Orange{ \sum_{\substack{M'_1 M'_2 M'_3 \\ M_p m_c m_d }} 
(-1)^{M'_2+m_c}  
\left(
  \begin{array}{ccc}
  1 & 1 & L_p \\
 -m_c & -m_d & M_p 
  \end{array}
 \right)
\left(
  \begin{array}{ccc}
  L'_1 & L'_2 & L'_3 \\
  M'_1 & - M'_2 & - M'_3 
  \end{array}
 \right)
\left(
  \begin{array}{ccc}
  L'_2 &  1 & L' \\
  M'_2 & -m_c & M' 
  \end{array}
 \right)
\left(
  \begin{array}{ccc}
  L'_3 & 1 & L \\
  - M'_3 & m_d & M 
  \end{array}
 \right) 
\left(
  \begin{array}{ccc}
  L'_1 & \ell_2 & L_p \\
  -M'_1 & m_2 & M_p 
  \end{array}
 \right) } \\
&\qquad\qquad = (-1)^{\Cyan{M'} + \ell_2 + L'_2 + L + 1 + L_p} 
\Cyan{ \left(
  \begin{array}{ccc}
  L' & L & \ell_2 \\
  M' & -M & m_2 
  \end{array}
 \right) } 
\left\{
  \begin{array}{ccc}
  L' & L & \ell_2 \\
  L'_2 & L'_3 & L'_1 \\
  1 & 1 & L_p 
  \end{array}
 \right\} ~,
\end{split}
\end{eqnarray}
and sum up the Wigner-$3j$ symbols over $M, M'$:
\begin{eqnarray}
\Cyan{ \sum_{M M'}(-1)^{M + M'}
\left(
  \begin{array}{ccc}
  L' & L & \ell_1 \\
  M' & -M & m_1 
  \end{array}
 \right) 
\left(
  \begin{array}{ccc}
  L' & L & \ell_2 \\
  M' & -M & m_2 
  \end{array}
 \right) } 
=  \frac{(-1)^{m_2}}{2 \ell_1 + 1} \delta_{\ell_1, \ell_2} 
\delta_{m_1, m_2} ~.
\end{eqnarray}
Following the same procedures in the other permutation and calculating
the summation over $L_p$ as
\begin{eqnarray} 
\sum_{L_p} I^{0 \lambda_2 -\lambda_2}_{L'_1 \ell_2 L_p} 
I^{0 \lambda_2 -\lambda_2}_{1 1 L_p} \frac{1 + (-1)^{L_p}}{2}
\left\{
  \begin{array}{ccc}
   L' & L & \ell_2\\
  L'_2 & L'_3 & L'_1 \\
  1 & 1 & L_p
  \end{array}
 \right\}
= - \frac{3}{2 \sqrt{2 \pi}} 
I_{L'_1 \ell_2 2}^{0 \lambda_2 -\lambda_2}
 \left\{
  \begin{array}{ccc}
  L' & L & \ell_2 \\
  L'_3 & L'_2 & L'_1 \\
  1 & 1 & 2
  \end{array}
 \right\}
~, 
\end{eqnarray}
we can obtain
the exact solution of Eq. (\ref{eq:def_pilm_power}) as 
 \begin{eqnarray}
\Braket{\Pi^{(\lambda_1)}_{Bv, \ell_1 m_1}(k_1)
 \Pi^{(\lambda_2) *}_{Bv, \ell_2 m_2}(k_2)} &=&
 - \frac{\sqrt{2\pi}}{3} 
\left(\frac{8 (2\pi)^{1/2}}{3 \rho_{\gamma,0}}\right)^2 
 / (2\ell_1 + 1) 
\delta_{\ell_1, \ell_2} \delta_{m_1,m_2}
\nonumber \\ 
&& \times \sum _{L L'}  \sum_{\substack{L_1 L_2 L_3 \\ L'_1 L'_2 L'_3}}
 (-1)^{\sum_{i=1}^3 \frac{L_i + L'_i}{2} }
I^{0~0~0}_{L_1 L_2 L_3} I^{0~0~0}_{L'_1 L'_2 L'_3}
\nonumber \\
&& \times 
\sum_{L_k} (-1)^{L'_2 + L_3} 
\left\{
  \begin{array}{ccc}
  L' & L & \ell_1 \\
  L_3 & L_2 & L_1 \\
  1 & 1 & L_k
  \end{array}
 \right\}
\left\{
  \begin{array}{ccc}
   L' & L & \ell_2\\
  L'_2 & L'_3 & L'_1 \\
  1 & 1 & 2
  \end{array}
 \right\} \nonumber \\
&& \times 
\Blue{ \int_0^\infty A^2 dA 
j_{L_1} (k_1 A)  
\int_0^\infty B^2 dB  
j_{L'_1} (k_2 B)  \nonumber \\
&& \times
\int_0^{k_D} k_1'^2 dk_1' {P}_B(k_1') j_{L_2} (k_1'A) j_{L'_3} (k_1' B)
\int_0^{k_D} k_2'^2 dk_2' {P}_B(k_2') j_{L'_2} (k_2'B) j_{L_3} (k_2' A) }
 \nonumber \\ 
&& \times   
\Red{ \sum_{S,S' = \pm 1} (-1)^{L_2 + L_2' + L_3 + L'_3} I^{0 S -S}_{L'_3 1 L} I^{0 S -S}_{L_2 1 L} I^{0 S' -S'}_{L_3 1 L'} 
I^{0 S' -S'}_{L'_2 1 L'} } \nonumber \\
&&  \times 
\OliveGreen{
\lambda_1 \lambda_2 
I^{0 \lambda_1 -\lambda_1}_{L_1 \ell_1 L_k} I^{0 \lambda_1
-\lambda_1}_{1 1 L_k} I^{0 \lambda_2 -\lambda_2}_{L'_1 \ell_2 2} } ~.
\end{eqnarray}
Note that in this equation, the dependence on the azimuthal quantum
number is included only in $\delta_{m_1, m_2}$. In the similar discussion of
the CMB bispectrum, this implies that the
CMB vector-mode power spectrum generated from the magnetized anisotropic
stresses is rotationally-invariant if the PMFs satisfy the statistical
isotropy as Eq. (\ref{eq:def_power}).

Furthermore, using such evaluations as
\begin{eqnarray}
&& \Red{ \sum_{S, S' = \pm 1} (-1)^{L_2 + L'_2 + L_3 + L'_3}
I^{0 S -S}_{L'_3 1 L} I^{0 S -S}_{L_2 1 L} 
I^{0 S' -S'}_{L_3 1 L'} I^{0 S' -S'}_{L'_2 1 L'} } \nonumber \\
&&\qquad\quad =  
\begin{cases}
4 I_{L'_3 1 L}^{0 1 -1} I_{L_2 1 L}^{0 1 -1} 
I_{L_3 1 L'}^{0 1 -1} I_{L'_2 1 L'}^{0 1 -1} 
& ({\rm for \ } L_3' + L_2, L_3 + L_2' = {\rm even} ) \\
0 & ({\rm otherwise})
\end{cases}~, \\
&& \OliveGreen{ \sum_{\lambda_1, \lambda_2 = \pm 1} 
I^{0 \lambda_1 -\lambda_1}_{L_1 \ell_1 L_k} I^{0 \lambda_1
-\lambda_1}_{1 1 L_k} I^{0 \lambda_2 -\lambda_2}_{L'_1 \ell_2 2} }
=
\begin{cases}
4 I_{L_1 \ell_1 L_k}^{0 1 -1} I_{1 1 L_k}^{0 1 -1}
I_{L_1' \ell_2 2}^{0 1 -1}  &
({\rm for \ } L_1 + \ell_1, L_1' + \ell_2 = {\rm even} ) \\
0 & ({\rm otherwise}) 
\end{cases} ~, \\
&& 
\Blue{ \left[ \prod_{n=1}^2 4\pi \int_0^\infty {k_n^2 dk_n \over
       (2\pi)^3} \mathcal{T}^{(V)}_{I, \ell_n}(k_n) \right]
\int_0^\infty A^2 dA j_{L_1} (k_1 A)  
\int_0^\infty B^2 dB j_{L'_1} (k_2 B)  \nonumber \\
&& \qquad\qquad \times
\int_0^{k_D} k_1'^2 dk_1' {P}_B(k_1') j_{L_2} (k_1' A) j_{L'_3} (k_1' B)
\int_0^{k_D} k_2'^2 dk_2' {P}_B(k_2') j_{L'_2} (k_2' B) j_{L_3} (k_2' A) }
 \nonumber \\ 
&& \qquad \simeq  
\left[ \prod_{n=1}^2 4\pi \int_0^\infty {k_n^2 dk_n \over
       (2\pi)^3} \mathcal{T}^{(V)}_{I, \ell_n}(k_n) 
j_{\ell_n} (k_n (\tau_0-\tau_*)) \right] \nonumber \\
&& \qquad\qquad\qquad \times A_B^2 (\tau_0 - \tau_*)^4 \left( \frac{\tau_*}{5} \right)^2 
\mathcal{K}_{L_2 L'_3}^{-(n_B +1)}(\tau_0-\tau_*)
\mathcal{K}_{L'_2 L_3}^{-(n_B +1)}(\tau_0-\tau_*) ~,
\end{eqnarray}
the CMB angle-averaged power spectrum is formulated as
\begin{eqnarray}
C^{(V)}_{I,\ell}
&\simeq& - \frac{\sqrt{2\pi}}{3} \left( \frac{32 (2\pi)^{1/2}}{3\rho_{\gamma,0}} \right)^2 / (2 \ell + 1)
\left[
4\pi \int_0^\infty {k^2 dk \over (2\pi)^3}
\mathcal{T}^{(V)}_{I, \ell}(k) j_{\ell} (k (\tau_0-\tau_*))
\right]^2 \nonumber \\
&& \times
\sum_{L_1 L_1'} 
\sum_{L_k} I_{L_1 \ell L_k}^{0 1 -1} I_{1 1 L_k}^{0 1 -1}
I_{L_1' \ell 2}^{0 1 -1} 
\sum_{L L'} 
\sum_{\substack{L_2 L'_2 \\ L'_3 L_3}} 
A_B^2 (\tau_0-\tau_*)^4 \left( {\tau_* \over 5} \right)^2 
\mathcal{K}_{L_2 L'_3}^{-(n_B +1)}(\tau_0-\tau_*)
\mathcal{K}_{L'_2 L_3}^{-(n_B +1)}(\tau_0-\tau_*) \nonumber \\
&& \times (-1)^{\sum_{i=1}^3 
\frac{L_i + L'_i}{2} + L'_2 + L_3} I^{0~0~0}_{L_1 L_2 L_3} 
 I^{0~0~0}_{L'_1 L'_2 L'_3} 
I_{L'_3 1 L}^{0 1 -1} I_{L_2 1 L}^{0 1 -1} 
I_{L_3 1 L'}^{0 1 -1} I_{L'_2 1 L'}^{0 1 -1}
\left\{
  \begin{array}{ccc}
  L' & L & \ell \\
  L_3 & L_2 & L_1 \\
  1 & 1 & L_k
  \end{array}
 \right\}
\left\{
  \begin{array}{ccc}
  L' & L & \ell \\
  L'_2 & L'_3 & L'_1 \\
  1 & 1 & 2
  \end{array}
 \right\}~. \label{eq:cmb_power} 
\end{eqnarray}
This has nonzero value in the configurations:
\begin{eqnarray}
\begin{split}
& (L_k, L_1) = (2, |\ell \pm 2|), (2, \ell), (1, \ell) ~, \ \ 
L'_1 = |\ell \pm 2|, \ell ~, \\
& |\ell - L| \leq L' \leq \ell + L ~, \\
& (L_2, L'_3) = (|L-1|,|L \pm 1|), (L,L), (L+1, |L \pm 1|) ~, \\
& (L'_2, L_3) = (|L'-1|,|L' \pm 1|), (L',L'), (L'+1, |L' \pm 1|)~, \\
& L_1 + L_2 + L_3 = {\rm even} ~, \ \ L_1' + L_2' + L_3' = {\rm even} ~, \\ 
& |L_1 - L_2| \leq L_3 \leq L_1 + L_2 ~, \ \ 
|L'_1 - L'_2| \leq L'_3 \leq L'_1 + L'_2 ~.
\end{split}
\end{eqnarray}

\begin{figure}[t]
  \begin{center}
    \includegraphics{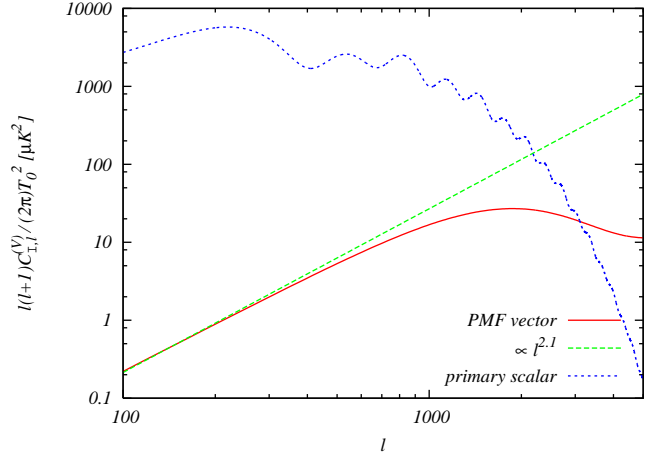}
  \end{center}
  \caption{(color online). The CMB power spectra of temperature
 fluctuation. The lines
 correspond to the spectra generated from vector anisotropic stress of PMFs as Eq.~(\ref{eq:cmb_power}) (red solid line)
 and primordial curvature perturbations (blue dotted line). The green dashed line express the asymptotic power of the red solid one.
The PMF parameters are fixed to $n_B = -2.9$ and $B_{1 {\rm Mpc}} = 4.7 {\rm nG}$, and the other cosmological parameters are fixed to the mean values limited from WMAP-7yr data reported in Ref. \cite{Komatsu:2010fb}.
  }
  \label{fig:vec_II_check}
\end{figure}

This shape is described in Fig.~\ref{fig:vec_II_check}. 
From this figure, we confirm that the amplitude and the overall behavior of the red solid line are in broad agreement with the previous studies (e.g. \cite{Lewis:2004ef, Yamazaki:2006bq, Paoletti:2008ck, Shaw:2009nf}). 
For $\ell \lesssim 2000$, using the scaling relations of the Wigner
symbols at the dominant configuration $L \sim \ell, L' \sim 1$ as discussed in Sec. \ref{sec:result}, we analytically find that $C^{(V)}_{I,\ell} \propto \ell^{n_B+3}$. This traces our numerical results as shown by the green dashed line.

\bibliography{paper}
\end{document}